\documentclass[journal]{IEEEtran}
\usepackage{amsmath,amsfonts}
\usepackage{algorithmic, soul}
\usepackage{algorithm}
\usepackage{listings}
\usepackage{array}
\usepackage[caption=false,font=normalsize,labelfont=sf,textfont=sf]{subfig}
\usepackage{hyperref}
\usepackage[nonumberlist]{glossaries}
\newacronym{3GPP}{3GPP}{3rd Generation Partnership Project}
\newacronym{5GB}{5GB}{fifth-generation of cellular networks and beyond}
\newacronym{5GS}{5GS}{5G system}
\newacronym{OS}{OS}{Operating System}
\newacronym{UE}{UE}{User Equipment}
\newacronym{RANs}{RANs}{Radio Access Networks}
\newacronym{RAN}{RAN}{Radio Access Network}
\newacronym{5GC}{5GC}{5G Core Network}
\newacronym{CN}{CN}{Core Network}
\newacronym{MNOs}{MNOs}{Mobile Network Operators}
\newacronym{SDRs}{SDRs}{Software-defined radios}
\newacronym{OAI}{OAI}{OpenAirInterface}
\newacronym{5G-SA}{5G-SA}{5G stand-alone}
\newacronym{TDD}{TDD}{Time-Division Duplexing}
\newacronym{FDD}{FDD}{Frequency-Division Duplexing}
\newacronym{DN}{DN}{Data Network}
\newacronym{CU}{CU}{Central Unit}
\newacronym{DU}{DU}{Distributed Unit}
\newacronym{O-DU}{O-DU}{O-RAN Distributed Unit}
\newacronym{RU}{RU}{Radio Unit}
\newacronym{O-RU}{O-RU}{O-RAN Radio Unit}
\newacronym{PHY-Low}{PHY-Low}{Lower Physical Layer}
\newacronym{RF}{RF}{Radio Frequency}
\newacronym{COTS}{COST}{Commercial Off-The-Shelf}
\newacronym{SDR}{SDR}{Software-Defined Radio}
\newacronym{FFT}{FFT}{Fats Fourier Transformation}
\newacronym{PTP}{PTP}{Precision Time Protocol}
\newacronym{SyncE}{SyncE}{Synchronous Ethernet}
\newacronym{USRP}{USRP}{Universal Software Radio Peripheral}
\newacronym{NFs}{NFs}{Network Functions}
\newacronym{E2E}{E2E}{end-to-end}
\newacronym{AMF}{AMF}{Access and Mobility Management Function}
\newacronym{SMF}{SMF}{Session Management Function}
\newacronym{UPF}{UPF}{User Plane Function}
\newacronym{UDM}{UDM}{Unified Data Management}
\newacronym{AUSF}{AUSF}{Authentication Server Function}
\newacronym{UDR}{UDR}{Unified Data Repository}
\newacronym{NRF}{NRF}{NF Repository Function}
\newacronym{PDU}{PDU}{Packet Data Unit}
\newacronym{NAS}{NAS}{Non Access Stratum}
\newacronym{PLMN}{PLMN}{Public Land Mobile Network}
\newacronym{5G-NSA}{5G-NSA}{5G-Non Standalone}
\newacronym{LTE}{LTE}{Long Term Evolution}
\newacronym{PDCP}{PDCP}{Packet Data Convergence Protocol}
\newacronym{RLC}{RLC}{Radio Link Control}
\newacronym{DL}{DL}{DownLink}
\newacronym{UL}{UL}{UpLink}
\newacronym{MCS}{MCS}{Modulation and Coding Scheme}
\newacronym{FR1}{FR1}{Frequency Range 1}
\newacronym{FR2}{FR2}{Frequency Range 2}
\newacronym{AGPLv3}{AGPLv3}{GNU Affero General Public License version 3}
\newacronym{SRS}{SRS}{Sounding Reference Signal}
\newacronym{UHD}{UHD}{USRP Hardware Driver}
\newacronym{BW}{BW}{Bandwidth}
\newacronym{GPSDO}{GPSDO}{GPS Disciplined Oscillator}
\newacronym{QSFP28}{QSFP28}{Quad Small Form-factor Pluggable 28}
\newacronym{GbE}{GbE}{Gigabit Ethernet}
\newacronym{EPC}{EPC}{Evolved Packet Core}
\newacronym{PCF}{PCF}{Policy and Charging Function}
\newacronym{NSSF}{NSSF}{Network Slice Selection Function}
\newacronym{BSF}{BSF}{Binding Support Function}
\newacronym{SEPP}{SEPP}{Security Edge Protection Proxy}
\newacronym{SCP}{SCP}{Service Communication Proxy}
\newacronym{QAM}{QAM}{Quadrature Amplitude Modulation}
\newacronym{RSRP}{RSRP}{Reference Signal Received Power}

\hypersetup{
    plainpages=false,       
    unicode=false,          
    pdftoolbar=true,        
    pdfmenubar=true,        
    pdffitwindow=false,     
    pdfstartview={FitH},    
    pdfnewwindow=true,      
    colorlinks=true,        
    linkcolor=black,         
    citecolor=black,        
    filecolor=magenta,      
    urlcolor=cyan           
}

\hypersetup{	
    colorlinks=true,%
    citecolor=black,%
    filecolor=black,%
    linkcolor=black,%
    urlcolor=black}
\makeglossaries

\usepackage{hhline}
\usepackage{textcomp}
\usepackage{caption}
\usepackage{stfloats}
\usepackage{soul}
\usepackage{setspace}
\usepackage{enumitem}
\usepackage{booktabs}
\usepackage{multirow}
\usepackage{bm}
\usepackage{url}
\usepackage{verbatim}
\usepackage{mathtools}
\usepackage{graphicx}
\usepackage[table,xcdraw]{xcolor}
\usepackage[
backend=biber,
maxnames=10,
]{biblatex}
\usepackage{tikz}
\usepackage{xcolor}
\usepackage{comment}
\usepackage{standalone}
\usetikzlibrary{positioning,shapes,shadows,arrows.meta}
\usepackage{verbatim}
\usepackage{pdflscape}
\newcommand{\red}[1]{\textcolor{red}{#1}}

\ifCLASSINFOpdf
\else
\fi
\begin{document}

%
\title{Performance Analysis and Comparison of Full-Fledged 5G Standalone Experimental TDD Testbeds in Single \& Multi-UE Scenarios}
%
%
%

\author{Maryam~Amini,
        and Catherine~Rosenberg,~\IEEEmembership{Fellow,~IEEE,}
        }

%
%


\newcommand{\ieeeheader}{\parbox{0.95\textwidth}{\small This work has been submitted to the IEEE for possible publication. Copyright may be transferred without notice, after which this version may no longer be accessible.}}

\markboth{\ieeeheader}
{\ieeeheader \hfill Shell \MakeLowercase{\textit{et al.}}: Bare Demo of IEEEtran.cls for IEEE Journals}

%



\maketitle


\begin{abstract}
Open-source software and Commercial Off-The-Shelf hardware are finally paving their way into the 5G world, resulting in a proliferation of  experimental 5G testbeds. Surprisingly, very few studies have been published on the comparative analysis of  testbeds with different hardware and software elements.

In this paper, we first introduce a precise nomenclature to characterize a 5G-standalone single-cell testbed based on its constituent elements and main configuration parameters. We then build 30 distinct such testbeds and systematically analyze their  performance with an emphasis on element interoperability (by considering different combinations of hardware and software elements from different sources), the number and  type of User Equipment (UE) as well as the Radio Access Network hardware and software elements  to address the following questions: 1) How is the  performance (in terms of bit rate and latency) impacted by different elements? 2) How does the number of UEs affect these results? 3) What is the impact of the user(s)' location(s) on the performance? 4) What is the impact of the UE type on these results?  5) How far does each testbed provide coverage? 6) And finally, what is the effect of the computing resources available to each open-source software? This study focuses on TDD testbeds.

\end{abstract}


%
\IEEEpeerreviewmaketitle

\section{Introduction}

\IEEEPARstart{A} fundamental shift in the architecture of mobile networks has happened, spearheaded by the \gls{3GPP}, which has introduced a novel, dis-aggregated, and open architecture for the  fifth-generation (5G) of cellular networks to enable \gls{MNOs} to source solutions tailored to their specific needs from various vendors.
The main elements 
of a 5G system remain the \gls{RAN}, the \gls{5GC}, as well as the \gls{UE}, but all of them  have gone through significant transformations. The advent of the NG-RAN~\cite{TS38.401}, 
a dis-aggregated architecture  composed of several vendor-neutral elements connected by open, standardized interfaces  with a major shift towards softwarization, marks a pivotal shift for cellular networks. Concurrently, the core network has adapted to host a multitude of novel \gls{NFs} designed to accommodate the diverse services envisioned for 5G. On the \gls{UE} side, a proliferation of devices with significantly differing features  creates new challenges for the network. 

In this context,  experimental testbeds have become crucial for test and validation and to verify interoperability, and pinpoint any gaps in the design of the different elements of this open, and dis-aggregated architecture.   
These testbeds benefit from the latest developments of: 
\begin{itemize}[wide, labelwidth=!, labelindent=0pt]
    \item  Software-defined radios (SDRs) that are hardware devices that serve as the radio component of 5G testbeds. Without them, over-the-air transmissions would need to be simulated, which would defeat the whole purpose of implementing a full-fledged experimental testbed.
    \item Several sophisticated software platforms and tools that have emerged both for the \gls{5GC} and the NG-RAN. Specifically, open-source frameworks, such as srsRAN~\cite{srsRAN}, which focuses on the \gls{RAN}, and \gls{OAI}~\cite{OAI}, which offers both  \gls{RAN} and  \gls{5GC} components\footnote{Both these platforms also offer 4G solutions.}, have gained significant momentum.
\end{itemize} 
This paper only focuses on  \gls{5G-SA} single cell testbeds that uses \gls{TDD}. We consider and build 30 such testbeds, varying in the software/hardware element of the \gls{RAN} as well as the number, and the type of the connected \gls{UE}s. Note that we have kept the \gls{5GC} the same in all those testbeds because we have shown in a previous paper~\cite{amini2024comparative} that performance is not really impacted by the core and the different \gls{RAN} elements do interoperate well with the \gls{5GC} that we have tried.

Most of the papers on \gls{5G-SA} open testbeds focus on studying the performance of single-\gls{UE} scenarios in terms of bit rate and latency, with occasional consideration given to coverage. Very few address interoperability and, to the best of our knowledge, no one addresses multi-UE scenarios conducted in multiple locations to examine the impact of location on the  performance of different types of \gls{UE}s.
A comprehensive study of multi-UE \gls{5G-SA} testbeds is yet to be done to fully unveil the potential of these experimental platforms and the interoperability of the different elements. 

This paper aims at shedding light on the impact of each element on the overall performance of a testbed in the context of multi-\gls{UE} scenarios. It examines how the location and the type of each UE plays a role in performance.
It also studies the interoperability of different types of UEs with different hardware and software elements of the RAN.
This paper synthesize and extensively expand our preliminary works reported in ~\cite{amini2023implementing, amini20235G, amini2024comparative}.  
Specifically, we have:
\begin{itemize}[wide, labelwidth=!, labelindent=0pt]
    \item Built and studied 28 single-cell \gls{5G-SA} \gls{TDD} testbeds, each differing by the combination of \gls{RAN} elements (software and \gls{SDR}) and  number and types of \gls{UE}(s) being used. We evaluated these testbeds from an interoperability perspective as well as from a performance standpoint, using well-defined quantitative and qualitative metrics, including data rate, latency, and coverage.
    \item Explored the multi-\gls{UE} case for different locations systematically, starting from good locations (please see Sec.~\ref{sec:Results} where we explain what we mean by ``good'') and progressively making the locations worse.
    \item Built two additional testbeds to evaluate the computational resource consumption of each software platform as the number of connected \gls{UE}s increases, by changing the PC on which the software platforms are hosted, offering a nuanced perspective on their strengths and limitations. This analysis aids researchers and practitioners in making informed decisions when selecting the appropriate software platforms and their host computing nodes, for their specific use cases. 
\end{itemize}


The rest of the paper is structured as follows: In Section~\ref{sec:Background}, we give the necessary background and present our nomenclature.
Section~\ref{sec:Literature} provides a comprehensive review of the relevant literature. In Section~\ref{sec:Platforms}, we introduce the different \gls{5G-SA} elements that we will consider.  Section~\ref{sec:Results} presents the metrics  used for our assessments, followed by the description,  methodology and results for each test scenario.
Section~\ref{sec:Conclusion} concludes the paper. An acronym table is given at the end of the paper.

\section{Background and Nomenclature}~\label{sec:Background}
In this section, we present the background material as well as the nomenclature used in the paper to fully characterize a single cell \gls{5G-SA} testbed. As mentioned earlier, the three primary elements of any cellular network are the core network, the \gls{RAN}, and the \gls{UE}. \gls{UE}s are devices that can be very different in terms of characteristics but they are all  equipped with a SIM card and seek connection to the cellular network. The \gls{RAN} provides access to the wireless medium to facilitate communication between the UEs and the core network. Finally, the core network is where all service and management aspects are handled. It also serves as the hub to connect the UEs to any external data network, including the Internet.


The migration from \gls{LTE}  to 5G is not  straightforward, as both the \gls{RAN} and core have been significantly changed. Indeed, all \gls{LTE} base stations and most of the core network should be replaced for a cellular network to be full-fledged  5G. Consequently, \gls{MNOs} have opted to transition to 5G in two phases, first  from \gls{LTE} to \gls{5G-NSA} and then from \gls{5G-NSA} to \gls{5G-SA}. In \gls{5G-NSA}, the core and control plane are \gls{LTE}-based, while the data plane follows  5G standards. This enables \gls{MNOs} to integrate 5G base stations into their existing \gls{LTE} network to handle the data plane and gradually transition their networks to a complete \gls{5G-SA} setup. 

In a  5G-SA system, the \gls{RAN}, which used to be a monolithic black box in \gls{LTE}, has now an open architecture with well-defined sub-elements and interfaces. 
Apart from the \gls{RF} element which is hardware-based, all other RAN elements are software-based and can be integrated and executed on a \gls{COTS}  computer. Similarly, the \gls{5GC} is characterized by a set of functionalities that are software based and can be executed in  \gls{COTS}  computers. Similarly, a UE can be decomposed into a software and a hardware element.

Thus, we can define a  single-cell, \gls{5G-SA} experimental testbed containing $n$ UEs, operating over-the-air, by the set $\mathcal{T}$ of its software and hardware elements (please see~(\ref{eq:Testbed})) and a  set $\mathcal{C}$ containing  the configuration parameters (please see~(\ref{eq:Config})).




\begin{equation} \label{eq:Testbed}
\begin{multlined}
    \mathcal{T} = \{S_{5GC}, H_{5GC}, S_{RAN}, H_{RAN}, \\
    (S_{UE_1}, H_{UE_1}), \dots, (S_{UE_n}, H_{UE_n})\}
\end{multlined} \qquad
\end{equation}

\begin{equation} \label{eq:Config}
\begin{multlined}
    \mathcal{C} = \{b, B\}
\end{multlined}
\end{equation}

\begin{figure}[ht]
    \centering
    \vspace{-6pt}
    \includegraphics[width=0.85\columnwidth]{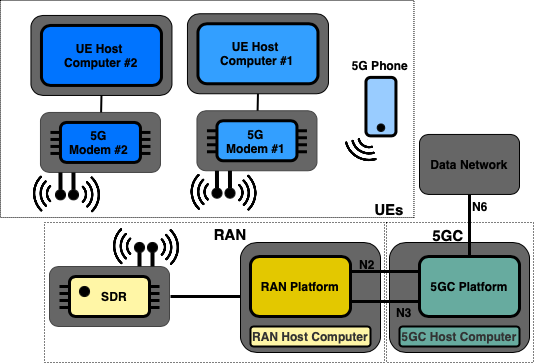}
    \caption{A typical single cell 5G-SA testbed}
    \label{fig:TestbedDesign}
    \vspace{-6pt}
\end{figure}

$S_{5GC}$ (resp. $S_{RAN}$ and $S_{UE_i}$) is the collection of  software  sub-elements for \gls{5GC} (resp. \gls{RAN} and the i-th \gls{UE}), and $H_{5GC}$ (resp. $H_{RAN}$ and $H_{UE_i}$) is the collection of hardware sub-elements for \gls{5GC} (resp. \gls{RAN} and the i-th \gls{UE}).
$b$ is the band central frequency, and $B$ is the bandwidth.
Kindly note that, the value of $b$ has a one to one mapping to the duplexing mode, i.e., \gls{TDD} or \gls{FDD}.
Fig.~\ref{fig:TestbedDesign} shows  the different elements of a \gls{5G-SA} testbed.
Starting with the \gls{RAN}, we can decompose it into four sub-elements: two hardware ones: i) a \gls{SDR} equipped with an antenna system responsible for the \gls{RF} front-end, ii) a computer to host the RAN software platform; and two software sub-elements: iii) a software platform to run the remaining 5G protocol stack,  and iv)  the Operating system (OS) of the computer. Clearly, a critical part of the RAN is the SDR.
In recent years thanks to the increased availability of \gls{SDR} devices, fast and cheap implementation of experimental 5G testbeds has become possible. 
Currently, the three major SDR vendors are: \textit{Ettus Research}~\cite{Ettus}, \textit{Lime Microsystems}~\cite{Lime}, and \textit{Nuand}~\cite{Nuand}. 
The two predominant open-source solutions for the $S_{RAN}$, are from \textit{srsRAN}~\cite{srsRAN} and \textit{OAI}~\cite{OAI}. 
As illustrated in Fig.~\ref{fig:TestbedDesign}, the \gls{SDR} gets connected to the computer hosting $S_{RAN}$ through a wired connection.
Then, the \gls{SDR} will exchange the radio samples with the \gls{RAN} software platform using a driver installed on the host computer.

In the past couple of years, \gls{5GC} solutions developed by \gls{OAI}~\cite{OAI}, \textit{Open5GS}~\cite{Open5GS} and \textit{free5GC}~\cite{free5GC}, have gained significant popularity among researchers. Each of these solutions supports different sets of \gls{5GC} \gls{NFs}. However, they all contain the essential \gls{NFs} required to implement an \gls{E2E} experimental testbed with basic functionality. These necessary \gls{NFs} are: 
Access and Mobility Management Function (AMF), Session Management Function (SMF), User Plane Function (UPF), Unified Data Management (UDM), Unified Data Repository (UDR), Authentication Server Function (AUSF), and NF Repository Function (NRF), ensuring \gls{UE} registration, authentication, Packet Data Unit (PDU) session establishment and management, and Non Access Stratum (NAS) security.
Also, much like the $S_{RAN}$, the $S_{5GC}$ needs a host computer and its OS for execution. 


Lastly, on the \gls{UE} front, there are three possible options. The most obvious one is to use a phone. Unfortunately, often, phones that are \gls{5G-SA} compatible cannot associate with  experimental testbeds. Some common reasons for such behavior include a discrepancy between the set of \gls{5G-SA} bands supported by the phone and the bands supported by the $S_{RAN}$, and the phone's failure to detect specific \gls{PLMN} identifiers. We found a 5G phone that was able to work in \gls{5G-SA} mode with all testbeds. Please see later for its description.
The other two options use a computer to host some of the \gls{UE} protocol stack. In the computer-based UE options that we have used, the computer is connected to a 5G modem that acts both as an \gls{RF} front-end and as a host for the lower layer protocols (up to Layer 3). Another computer-based UE option is to use an SDR as an \gls{RF} front-end while the computer hosts $S_{UE}$ (both \textit{srsRAN} and \textit{OAI} offer \gls{UE}-based software platforms). As discussed in~\cite{amini2024comparative}, 5G modems have proven to be more convenient for testing than phones due to their support for multiple \gls{5G-SA} bands, their ability to associate with non-public networks, and their ease of configuration. For completeness, we note that another sub-element of the \gls{UE}  is the SIM card. The use of a programmable SIM card enables the modification of authentication information on the \gls{UE}, based on the testbed's requirements.


With respect to the testbed depicted in Fig.~\ref{fig:TestbedDesign}, please note that, often, the number of computers is reduced by putting multiple software platforms (e.g., $S_{RAN}$ and $S_{5GC}$) on the same computer. In this study, we will build, analyze and compare the performance, measured on the \gls{UE}-side, of different single-cell 5G SA testbeds using \gls{TDD}, i.e., made of different combinations of elements and sub-elements. 
The elements/sub-elements under-study are described in Section~\ref{sec:Platforms}. We will also  show how computational resources will affect the  performance of the testbed and how different software platforms utilize those resources. To keep this study tractable (in terms of number of testbeds) and due to page limitation, we have restricted ourselves to \gls{TDD}-based testbeds. A similar study on FDD testbeds is planned for future. 
\section{Literature Review}~\label{sec:Literature}
In this section, we provide a comprehensive overview of the existing literature related to experimental open 5G-SA testbeds. While there are a number of publications on this subject, our study deliberately narrows its focus to papers featuring experimental testbeds equipped with all essential elements for an operational functionality, as opposed to simulating or emulating parts of the testbed.

Next, we first review papers that have focused on a single full-fledged 5G-SA testbed and then those that dealt with  comparisons of full-fledged testbeds.


\subsection{Targeted Studies on the performance of a single testbed}
Haakegaard \emph{et al.} focus on  a 5G-SA testbed, 
operating in TDD mode, 
employing \textit{Open5GS} and \textit{srsRAN} in ~\cite{haakegaard2024performance}. This study compares the theoretical and achieved performance of the testbed in terms of \gls{UL} \& \gls{DL} bit rates, latency, and coverage.  Additionally, the authors study the effect of several radio parameters, such as the number of \gls{SDR} antennas, the bandwidth, and the  Time-Division Duplexing (TDD) frame structure, on the performance of the testbed.

In~\cite{bozis2024versatile}, Bozis \emph{et al.}  present their \gls{5G-SA} testbed, operating in TDD mode over band \textit{n78}
which features the \gls{5GC}, \gls{RAN}, and \gls{UE} solutions from the \gls{OAI} project. In this testbed, two \gls{USRP} N310 devices are utilized for the \gls{SDR}-based UE and the \gls{RAN}. The RAN and UE  are connected through \gls{RF} cables instead of wirelessly, over-the-air. This study reports the latency and \gls{DL}/\gls{UL} throughput of a single-UE for two bandwidth scenarios.

In the evaluation done by Chepkoech \emph{et al.} \cite{chepkoech2023evaluation}, the performance of six testbeds operating in \gls{LTE}, \gls{5G-NSA}, and \gls{5G-SA} modes are studied with a focus on metrics such as throughput, latency, and signal strength. Notably, the only testbed operating in \gls{5G-SA} mode is implemented using \textit{srsRAN\_4G}, and \textit{Open5GS}, and over band \textit{n7}, which is an FDD 5G-NR band.

Sahbafard \emph{et al.}  provide a comprehensive assessment of a \gls{5G-SA} testbed, 
operating on TDD mode
and utilizing \gls{OAI} for both the \gls{RAN} and the \gls{5GC} platforms in \cite{sahbafard2023performance}. This testbed, uses Quectel 5G modems as \gls{UE}s. The authors compare the modem's achieved performance while using \gls{USRP} B210 or N310 as the \gls{SDR} of choice. They also conduct an analysis of the signal strength, in both single-user and multi-user scenarios, to evaluate the testbed's coverage. 

The authors of~\cite{mamushiane2023deploying} provide a tutorial on establishing a slice-aware \gls{5G-SA} testbed utilizing \textit{srsRAN\_Project} and \textit{Open5GS} for the \gls{RAN} and \gls{5GC} software platforms. The paper provides insights into the challenges of integration of different elements to the testbed. Furthermore, it offers valuable information on potential issues faced during the implementation phase, along with troubleshooting strategies for these scenarios. Note that, the authors have not mentioned what specific band they are using for their tests. They are using 30 KHz for subcarrier spacing which is mostly used for TDD bands.

\subsection{Comparative studies on the performance of multiple testbeds}
The literature addressing  comparative analysis of 5G-SA testbeds with different combinations of elements is quite scarce. While the inherent design of open-source software platforms aims to facilitate interoperability, it is crucial to verify its feasibility, simplicity, and performance. 
To the best of our knowledge, the only existing studies on this subject are~\cite{alvesexperimental,  chepkoech2023oss, gabilondo20215g, mubasier2023campus} as well as our conference papers~\cite{amini20235G, amini2024comparative}.

The authors of~\cite{alvesexperimental} have provided a comparison between the performance of \textit{srsRAN} and \textit{\gls{OAI}} in three aspects, namely, \gls{UE}'s \gls{DL} bit rate, latency, and a qualitative comparison of the quality of a video call made by the \gls{UE}. More so, they assessed the interoperability of the employed open-source \gls{RAN} and \gls{5GC} software.
The authors have configured all their testbeds to be operating over band \textit{n78}, in TDD mode.
Additionally, the study highlights the  differences in performance between \gls{SDR}-based \gls{UE}s and \gls{COTS} \gls{UE}s. 
This study links the differences between the rates achieved by \gls{OAI} and those achieved by srsRAN to the differences in their Quadrature Amplitude Modulation (QAM) implementation. 
Later, in Section~\ref{sec:Results}, we will show that another explanation might be linked to the fact that the computational resources available to $S_{RAN}$ have a significant impact on the UE's achieved rate.


\cite{gabilondo20215g} presents a comprehensive assessment of the performance of three \gls{5GC} software: \textit{Open5GS, Open5GCore, and Amarisoft 5G Core} for three types of 5G modems, in terms of both throughput and latency, when the same \gls{RAN}, \textit{Amarisoft 5G RAN} is used. 

In~\cite{mubasier2023campus}, Mubasier \emph{et al.} have implemented two distinct testbeds. both operating on band \textit{n77}, which is a TDD band. 
The first testbed features \textit{\gls{OAI}} \gls{5GC} and \gls{RAN}, along with \gls{USRP} B210 and a host laptop. The second is a testbed utilizing \gls{USRP} X300 in conjunction with \textit{srsRAN} and \textit{Open5GS}. This study then evaluates network connectivity, the performance of the testbed seen by the \gls{UE}, and also computing resource utilization of the open-source software.

\cite{chepkoech2023oss} is another comparative study on the performance of testbeds comprising different elements. In this study  \textit{\gls{OAI}}, and \textit{srsRAN} were the focus, and a comparative analysis of their features, as well as quantitative results in terms of throughput, signal strength and latency were discussed for two \gls{5GC} software, namely, \textit{Open5GS}, and \textit{Free5GC}, in single-\gls{UE} scenario. 
All the tests were set to be conducted over band \textit{n78}.

In our previous study~\cite{amini2024comparative}, we focused on evaluating the interoperability of 5G open-source software by examining the \gls{UE}'s achieved performance for various combinations of software platforms in a \gls{5G-SA} testbed. Our tests were all done over band \textit{n78}.
We showed that the choice of \gls{5GC} does not affect the performance observed by the \gls{UE}.
Earlier in~\cite{amini20235G}, we compared two \gls{5G-SA} testbeds that were different only in $S_{RAN}$ for two different \gls{SDR} devices, namely, \gls{USRP} B210 \& X410. We also studied the effect of the connectivity mode between the \gls{SDR} and the \gls{UE}, i.e., wired or wireless. 
In this study, we selected band \textit{n3}, an FDD 5G-NR band, for our tests and comparisons.

This paper is a comparative study of several \gls{5G-SA} experimental testbeds. 
We have used \textit{Open5GS} as the fixed \gls{5GC} for all the testbeds. Moreover, we employed the same set of configuration parameters, i.e., the same frequency band and bandwidth, for all the testbeds to ease the comparison of the results. The analysis in this study is conducted from two perspectives: 
\begin{itemize}[wide, labelwidth=!, labelindent=0pt]
    \item The performance achieved by the \gls{UE}s. This is an extension of what we did in ~\cite{amini20235G, amini2024comparative}. The extension is on several fronts: we consider coverage as a new metric, different types of UEs, and  different locations, as well as multi-\gls{UE} scenarios. 
    \item The computational resource consumption of the different software elements, and the impact of their host PC on the performance.
\end{itemize}

\section{Elements \& Testbeds under-study}~\label{sec:Platforms}
\begin{figure}[h]
    \centering
    \includegraphics[width=0.95\columnwidth]{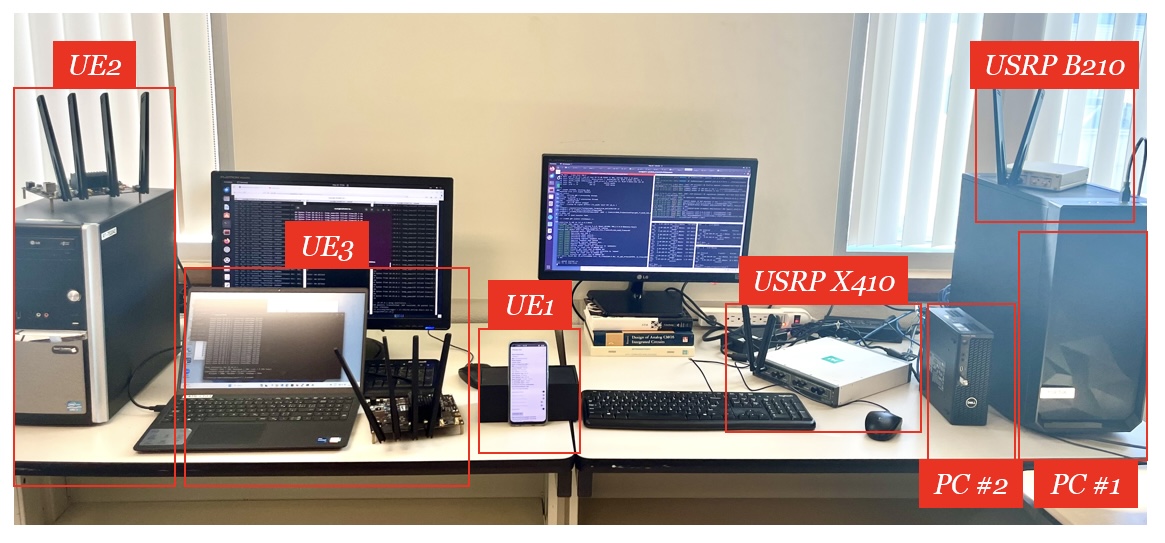}
    \caption{Our testbeds elements}
    \label{fig:Lab}
\end{figure}

In this section, we will provide a detailed description of the various elements and sub-elements that have been utilized for the \gls{5G-SA} testbeds of our comparative study. 
The list of those elements/sub-elements is given in Table~\ref{tab:PlatformElements}. 
The hardware elements of our testbeds are shown in Fig.~\ref{fig:Lab}.
We then introduce the testbeds that we have considered and built for this study. 
We have considered \emph{all} possible combinations of \gls{5GC}, \gls{RAN} and \gls{UE} sub-elements.

\begin{table*}[ht]
\centering
\caption{Our 5G-SA platform elements
\\ (We use the same color code as in Fig.~\ref{fig:TestbedDesign})} 
\label{tab:PlatformElements}
\begin{adjustbox}{width=0.8\textwidth}
\begin{tabular}{|cc|cc|}
\hline
\multicolumn{2}{|c|}{\cellcolor[HTML]{000000}{\color[HTML]{FFFFFF} \textbf{Software Elements}}}                                         & \multicolumn{2}{c|}{\cellcolor[HTML]{000000}{\color[HTML]{FFFFFF} \textbf{Hardware Elements}}}                                                  \\ \hline
\multicolumn{1}{|c|}{\cellcolor[HTML]{67AB9F}}                                        & Open5GS                      & \multicolumn{1}{c|}{\cellcolor[HTML]{67AB9F}}                                        &                                                          \\ \cline{2-2}
\multicolumn{1}{|c|}{\cellcolor[HTML]{67AB9F}}                                        & OS: Ubuntu 20.04.6 LTS                          & \multicolumn{1}{c|}{\cellcolor[HTML]{67AB9F}}                                        &                                                          \\ \cline{2-2}
\multicolumn{1}{|c|}{\multirow{-3}{*}{\cellcolor[HTML]{67AB9F}\textbf{$S_{5GC}$}}}   & Kernel: 5.15.0-xx-lowlatency                    & \multicolumn{1}{c|}{\multirow{-3}{*}{\cellcolor[HTML]{67AB9F}\textbf{$H_{5GC}$}}}   & \multirow{-3}{*}{\begin{tabular}[c]{@{}l@{}}5GC/RAN Host PC $\in$ \{Host Computer \#1 \& \#2\} \\ (both $S_{5GC}$ and $S_{RAN}$ run on the same host PC) \end{tabular} } \\  
\hline \hline
\multicolumn{1}{|c|}{\cellcolor[HTML]{E3C800}}                                        & $S_{RAN}$ $\in$ \{srsRAN, OAI\}                          & \multicolumn{1}{c|}{\cellcolor[HTML]{FFF6A6}}                                        &                                                          \\ \cline{2-2}
\multicolumn{1}{|c|}{\cellcolor[HTML]{E3C800}}                                        & OS: Ubuntu 20.04.6 LTS                          & \multicolumn{1}{c|}{\cellcolor[HTML]{FFF6A6}}                                        & \multirow{-2}{*}{5GC/RAN Host PC $\in$ \{Host Computer \#1 \& \#2\}} \\ \cline{2-2} \cline{4-4} 
\multicolumn{1}{|c|}{\cellcolor[HTML]{E3C800}}                                        & Kernel: 5.15.0-xx-lowlatency                    & \multicolumn{1}{c|}{\cellcolor[HTML]{FFF6A6}}                                        & SDR $\in$ \{USRP B210 \& USRP X410\}                         \\ \cline{2-2} \cline{4-4} 
\multicolumn{1}{|c|}{\multirow{-4}{*}{\cellcolor[HTML]{E3C800}\textbf{$S_{RAN}$}}}   & Driver: UHD v4.5.0                              & \multicolumn{1}{c|}{\multirow{-4}{*}{\cellcolor[HTML]{FFF6A6}\textbf{$H_{RAN}$}}}   & Antenna: Molex 5GHz compatible                           \\ \hline \hline
\multicolumn{1}{|c|}{\cellcolor[HTML]{99CCFF}}                                        &                                                 & \multicolumn{1}{c|}{\cellcolor[HTML]{99CCFF}}                                        & 5G Phone: OnePlus Nord CE 2                              \\ \cline{4-4} 
\multicolumn{1}{|c|}{\multirow{-2}{*}{\cellcolor[HTML]{99CCFF}\textbf{$S_{UE_1}$}}} & \multirow{-2}{*}{OS: Android 11}                & \multicolumn{1}{c|}{\multirow{-2}{*}{\cellcolor[HTML]{99CCFF}\textbf{$H_{UE_1}$}}} & SIM Card: sysmoISIM-SJA2                                 \\ \hline \hline
\multicolumn{1}{|c|}{\cellcolor[HTML]{339FFF}}                                        & OS: Ubuntu 20.04.6 LTS                          & \multicolumn{1}{c|}{\cellcolor[HTML]{339FFF}}                                        & Host PC: Intel® CoreTM i7-3770 CPU @ 3.4GHz              \\ \cline{2-2} \cline{4-4} 
\multicolumn{1}{|c|}{\cellcolor[HTML]{339FFF}}                                        &                                                 & \multicolumn{1}{c|}{\cellcolor[HTML]{339FFF}}                                        & 5G Modem: Quectel RM502Q-AE                              \\ \cline{4-4} 
\multicolumn{1}{|c|}{\multirow{-3}{*}{\cellcolor[HTML]{339FFF}\textbf{$S_{UE_2}$}}} & \multirow{-2}{*}{Kernel: 5.14.0-051400-generic} & \multicolumn{1}{c|}{\multirow{-3}{*}{\cellcolor[HTML]{339FFF}\textbf{$H_{UE_2}$}}} & SIM Card: sysmoISIM-SJA2                                 \\ \hline \hline
\multicolumn{1}{|c|}{\cellcolor[HTML]{0074FF}}                                        &                                                 & \multicolumn{1}{c|}{\cellcolor[HTML]{0074FF}}                                        & Host PC: Intel® Core i7-1255U CPU @ 1.70 GHz                  \\ \cline{4-4} 
\multicolumn{1}{|c|}{\cellcolor[HTML]{0074FF}}                                        &                                                 & \multicolumn{1}{c|}{\cellcolor[HTML]{0074FF}}                                        & 5G Modem: Quectel RM502Q-AE                              \\ \cline{4-4} 
\multicolumn{1}{|c|}{\multirow{-3}{*}{\cellcolor[HTML]{0074FF}\textbf{$S_{UE_3}$}}} & \multirow{-3}{*}{OS: Windows 11}                & \multicolumn{1}{c|}{\multirow{-3}{*}{\cellcolor[HTML]{0074FF}\textbf{$H_{UE_3}$}}} & SIM Card: sysmoISIM-SJA2                                 \\ \hline
\end{tabular}%
\end{adjustbox}
\end{table*}

\subsection{The RAN}
\subsubsection{RAN Software Platforms}
\begin{itemize}[wide, labelwidth=!, labelindent=0pt]
    \item \emph{Platform \#1- srsRAN}: It comprises open-source 4G and 5G software radio suites developed by the \textit{Software Radio Systems} team. The project includes two main repositories, namely, \textit{srsRAN\_4G} and \textit{srsRAN\_Project}, both available under the GNU Affero General Public License version 3 (AGPLv3). While \textit{srsRAN\_4G} provides a prototype for \gls{5G-SA}, the supported features are minimal, and there will be no further updates. \textit{srsRAN\_Project} though, offers a full \gls{5G-SA} solution based on a complete codebase. 
In this study we have used \textit{srsRAN\_Project}, and we will refer to it as \textit{srsRAN} in the following.
It 
supports all \gls{TDD} and \gls{FDD} bands on Frequency Range 1 (FR1). The latest release of the software (srsRAN\_Project 23.10.1) offers the flexibility to configure over 400 parameters in a user-friendly way, which has made working with \textit{srsRAN} particularly convenient. 
\item \emph{Platform \#2- OAI-RAN}: Developed by the \textit{Eurecom} team, the \gls{OAI} software platform provides \gls{LTE}, and 5G solutions for the \gls{RAN}, and unlike srsRAN, the 5G core. For clarity, we will refer to the \gls{RAN} solution from \gls{OAI}, as \textit{OAI-RAN}, and the \gls{5GC} as \textit{OAI-5GC}. 
This project is distributed under the \textit{OAI 5G Public License}.
Compared to \textit{srsRAN}, \textit{OAI-RAN} provides more features such as more subcarrier spacing options and  support for Frequency Range 2 (FR2). However, configuring OAI-RAN is more complex than configuring srsRAN since it requires modifying code blocks in the configuration file. 
In this regard, we highly recommend reading~\cite{seidel2023get}, where the authors have described how to work with the \textit{OAI-RAN} configuration file. 
We have used the \textit{``2024.w09''} version of the \textit{develop} branch of \textit{OAI}'s GitLab. 
\end{itemize}

\subsubsection{RAN Hardware Platforms (SDR)}
\begin{itemize}[wide, labelwidth=!, labelindent=0pt]
    \item \emph{RAN SDR \#1- USRP X410}: it is a high-end, all-in-one \gls{SDR}.
It comes with advanced features like four independent Tx and Rx channels, each capable of 400 MHz of bandwidth. The X410 model is equipped with a built-in GPS Disciplined Oscillator (GPSDO) for improved timing synchronization. Additionally, it offers multiple networking interfaces for data and control offloading, such as two Quad Small Form-factor Pluggable 28 (QSFP28) ports supporting data transfer rates of up to 100 Gigabit Ethernet (GbE), along with standard interfaces like Ethernet and USB-C. In our experiments, we utilize one \gls{USRP} X410 connected to the RAN host computer via two QSFP28-10GB connections and one Ethernet connection to the network.
 \item \emph{RAN SDR \#2- USRP B210}: This single-board, low-cost \gls{USRP} is a dual-channel transceiver, providing up to 56 MHz bandwidth.  B210 comes with a USB 3.0 connector to enable a connection to the RAN host PC. Since this \gls{USRP} lacks a built-in GPSDO, maintaining synchronization might become a challenge.
We have chosen to use a \gls{USRP} B210 in our tests, as it is arguably the most popular \gls{SDR} in the research community as of now. Hence, we can gain a clear understanding of what  this \gls{USRP} model offers compared to the high-end X410.
\end{itemize}

\subsubsection{RAN Host Computers}
To investigate the influence of computing resources on the testbed's performance, we utilize the two PCs listed in Table~\ref{tab:PlatformElements}, featuring different levels of computational power to host the RAN software. 

\begin{itemize}[wide, labelwidth=!, labelindent=0pt]
    \item \emph{PC \#1}: The first  host computer utilized in our tests is equipped with an $11^{th}\,Gen\,Intel^{(R)}\,Core^{TM}\,i9-11900K$ processor, running at the base frequency of 3.50GHz. This system operates on Ubuntu 20.04.6 LTS, featuring kernel version 5.15.0-60-low-latency.
  \item \emph{PC \#2}:
The second  host is a mini PC featuring $Intel^{(R)}\,Core^{(TM)}\,i7-10700$ CPU @ 2.90GHz. This PC also runs Ubuntu 20.04.6 LTS, with kernel version 5.15.0-84-low-latency.
\end{itemize}

Note that the \gls{SDR} requires a driver installed on the RAN host computer, so that they both can communicate. All \gls{USRP} products from \textit{Ettus} use the same hardware driver, called USRP Hardware Driver (UHD). In this study, we have installed UHD\_4.5.0.0 on both RAN host computers.
\subsection{The Core}
\subsubsection{5GC Software Platforms}
As discussed earlier, we are restricting the study to a single \gls{5GC} software platform, namely,
\emph{Open5GS}.
It is a popular core network solution that not only offers a \gls{5GC} but also an Evolved Packet Core (EPC) solution, enabling the implementation of \gls{5G-SA}, \gls{5G-NSA}, and \gls{LTE} networks. The \gls{5GC} solution is based on \gls{3GPP}-Rel.17, and contains the following network functions: NRF, Service Communication Proxy (SCP), Security Edge Protection Proxy (SEPP), AMF, SMF, UPF, AUSF, UDM, UDR, Policy and Charging Function (PCF), Network Slice Selection Function (NSSF), and Binding Support Function (BSF). It is open-source and available under AGPLv3. For our test scenarios, we have used \textit{Open5GS} v2.7.0.

\subsubsection{5GC Hardware Platforms}
The core hardware platform is one of the two PCs described above since we execute both  $S_{5GC}$ and $S_{RAN}$ on the same host computer.
\subsection{The UEs}
We consider three different UEs.
\begin{itemize}[wide, labelwidth=!, labelindent=0pt]
    \item \emph{UE$_1$}: 
    Our first \gls{UE} is a \textit{OnePlus Nord CE 2 5G} \gls{COTS} phone, which is \gls{5G-SA} compatible. This phone runs \textit{Android 11} and supports 11 \gls{5G-SA} bands. In order to force this phone to operate on \gls{5G-SA} mode only, we installed an Android application called \textit{5G Switch - Force 5G Only}~\cite{force5G}. This application is free and does not require the phone to be rooted. 
\item \emph{UE$_2$}:
The second \gls{UE}  comprises a Quectel 5G modem, the RM502Q-AE, connected to a host PC. 
The PC ($Intel^{(R)}\,Core^{TM}\,i7-3770$ CPU @ 3.4GHz) is running Ubuntu 20.04 with  kernel version  v5.14.0. For further details on the challenges encountered during the setup of this UE, please refer to~\cite{amini2023implementing}, where we have described the necessary configurations for this type of a \gls{UE}. Note that this UE is not easily movable.
\item \emph{UE$_3$}:
The third \gls{UE} is composed of another Quectel 5G modem (RM502Q-AE) connected to a Dell laptop equipped with a $12^{th}$ Generation Intel Core i7-1255U processor running at 1.70 GHz and Windows 11. This choice allows us to investigate potential performance differences between the second and third \gls{UE}s and determine if these differences can be attributed to their respective host computer operating systems (Ubuntu vs. Windows).
\end{itemize}

\subsection{Miscellaneous}
Last but not least, we utilized \textit{sysmoISIM-SJA2} programmable SIM cards from \textit{sysmocom}~\cite{sysmocom} in this study. These SIM cards are 3GPP-Rel.16 compliant and come with the credentials required for modifying them. Additionally, for configuring our testbed's \gls{PLMN}, we assigned the Mobile Country Code (\textit{MCC}), and Mobile Network Code (\textit{MNC}) values as \textit{001} and \textit{01}, respectively.

\subsection{Testbeds Under-Study}
Now that we have introduced all the elements under study, we can present the testbeds that we have built. With respect to the definitions of the sets $\mathcal{T}$ and $\mathcal{C}$ in ~(\ref{eq:Testbed}) and (\ref{eq:Config}) respectively.
We have configured all the tests to be done on $b=n78$, and $B=40$~MHz bandwidth, with sub-carrier spacing (SCS) equal to 30 kHz. Please note that, \textit{n78} operates in \gls{TDD} mode. Consequently, it is imperative to carefully configure the  same \gls{TDD} slots and symbols in both $S_{RAN}$ for a meaningful comparison. We set the frame structure to be: ``DDDDDDFUUU'', accounting for 6 \gls{DL} slots, 3 \gls{UL} slots, and 1 Flexible slot. Additionally, we picked \textit{PC \#1} as the host computer, running $S_{5GC}$ and $S_{RAN}$.
We have build and analyzed the performance of 28 testbeds that consist of \emph{all the possible combinations} of the other elements described above with either (any) one, (any) two or the three UEs.  
Specifically, considering the three UE devices that we have described above, we created seven different UE combinations based on their number and types, i.e., \{UE$_1$\}, \{UE$_2$\}, \{UE$_3$\}, \{UE$_1$, UE$_2$\}, \{UE$_1$, UE$_3$\}, \{UE$_2$, UE$_3$\}, \{UE$_1$, UE$_2$, UE$_3$\}. Recall that we consider two \gls{SDR} devices, i.e., \{\textit{USRP B210}, \textit{USRP X410}\} and two $S_{RAN}$ platforms, i.e., \{\textit{srsRAN}, \textit{OAI-RAN}\}. Thus, using all possible combinations of these three groups, (7 $\times$ 2 $\times$ 2), we built 28 testbeds.

Additionally,  to assess the computational resource consumption of the two 5G open-source software, we have built two additional testbeds using a less powerful PC, PC \#2, and the two \gls{RAN} platforms. This setup allows us to evaluate how the performance of each open-source software is impacted by the host PC, and the number of connected \gls{UE}s. Hence in total, we have build and studied 30 testbeds.

\section{Test Scenarios, Methodology and Results}~\label{sec:Results}

In this section, we first define the metrics we use to assess the performance of the \gls{5G-SA} testbeds.
We then introduce the test scenarios and the corresponding methodology, followed by a presentation of the results
\begin{figure*}[ht]
    \centering
    \includegraphics[width=0.75\textwidth]{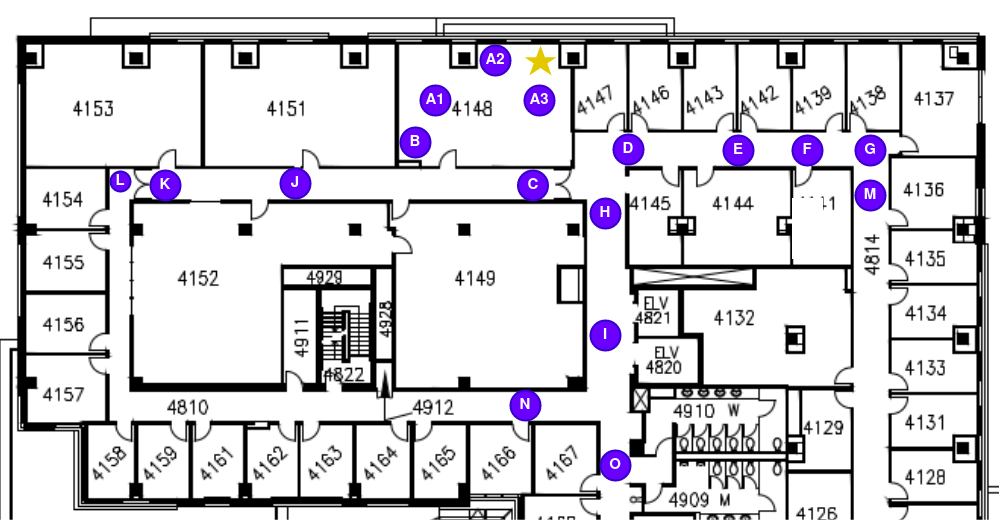}
    \caption{Map indicating the positions where tests were conducted}
    \label{fig:Coverage}
\end{figure*}

\subsection{Performance Metrics}

\subsubsection{Data Rate}
We measure the \gls{UL} and \gls{DL} average data rate in Mbps, using \textit{iperf3}~\cite{iperf3}. Each experiment runs for three minutes, and we report the average of the achieved \gls{UL} and \gls{DL} data rates. Please note that, to run \textit{iperf3} on UE$_1$ the Android phone, we installed \textit{he.net - Network Tools}~\cite{he.net}. 


\subsubsection{Latency}
The \gls{E2E} latency is measured in milli-second (ms),  by using the \textit{ping} command at the UE side. 
We conducted each test for three minutes and report the average latency between the \gls{UE} and the \gls{5GC}.



\subsubsection{Coverage}
We consider the \gls{RSRP} in dBm to be the measure of the coverage. The coverage tests were conducted for UE$_1$.
We used the Android application \textit{5G Switch - Force 5G Only}~\cite{force5G}, on UE$_1$, the mobile phone, to report the \gls{RSRP}.




\subsubsection{Computational resource consumption}
Finally, to monitor how each \gls{RAN} software platform consumes the computational resources of its host computer, we use the \textit{top} command. This allows us to observe the running processes, and the overall host computer resource utilization (CPU and memory).
We run this command on the host PC that executes both \gls{5GC} and \gls{RAN}, for three minutes and report the maximum percentage of CPU and memory utilization for each software platform.



\subsection{Methodology \& Results for Data Rate Assessment}
\subsubsection{Methodology}

For the assessment of the data rate of each UE within each testbed, we had to carefully take the location of each \gls{UE} into account. The first tests were done when all the \gls{UE}s of each testbed were located in \textit{``good''} positions, i.e., at positions where the downlink data rate of a single UE was consistently at its peak (characterized by the highest rate of the existing \gls{MCS}). The comparison of the performance of the different testbeds when all \gls{UE}s are in those positions, gives us valuable information on how best the testbeds can perform.
    
Fig.~\ref{fig:Coverage} illustrates the map of the fourth floor of the \textit{Centre for Environmental and Information Technology} building on the main campus of University of Waterloo where we conducted our tests. Our lab is in room \textit{4148}, and we have indicated the location of the \gls{SDR}  by a star sign on the map.
    
We first identified three  good positions in our lab, all in the vicinity of the \gls{SDR} device. We have marked the three selected positions for the \gls{UE}s in Fig.~\ref{fig:Coverage} as $A1, A2, A3$. UE$_1$ was placed at $A1$, UE$_2$  at $A2$, and UE$_3$ at $A3$. Recall that UE$_2$ is not easily movable and hence was kept at $A2$ for all tests.
Hence, for all other locations, we only checked the rates seen by UE$_1$ and UE$_3$. 
The results of this initial round of tests where the UEs are in good positions, with either one, two or three \gls{UE}s are presented in Table~\ref{tab:DataRate}  (please refer to  tests \{T1,T2,T3,T8,T9,T10,T15\}).
    
Next, to study the impact of locations on the  data rate observed by the \gls{UE}(s), we selected two additional positions where the drop in  the data rate was significant enough to categorize the positions as \textit{``fair''}, and \textit{``bad''}. In this regard, after multiple preliminary tests conducted using UE$_1$, and UE$_3$, position $D$ on the map was selected as the position which would yield a \textit{``fair''} rate, with \gls{MCS} values observed between 15 to 17 and position $E$ as the \textit{``bad''} position, with \gls{MCS} values observed between 9 to 11. The single and multi-UE results corresponding to these tests are presented in Table~\ref{tab:DataRate}, with test ids \{T4,T5,T6,T7,T11,T12,T13,T14,T16,T17\}. Note that the other positions in Fig.~\ref{fig:Coverage} are used for our coverage study.

In order to keep the number of tests reasonable and the size of the tables manageable, we only used 14 of the 28 testbeds to study the impact of location on performance, by fixing the \gls{SDR} to \gls{USRP} B210 in this first round of tests. A second round of tests to compare the SDR, is described later in the paper.
    

\begin{table}[]
\centering
\caption{Downlink and uplink data rate results for different tests corresponding to different locations and $S_{RAN}$ (in Mbps)}
\label{tab:DataRate}
\resizebox{0.95\columnwidth}{!}{%
\begin{tabular}{|ccccccc|}
\hline
\rowcolor[HTML]{329A9D} 
\multicolumn{1}{|c|}{\cellcolor[HTML]{329A9D}}                                             & \multicolumn{2}{c|}{\cellcolor[HTML]{329A9D}{\color[HTML]{000000} \textit{$S_{RAN}$}}}                                                                                          & \multicolumn{2}{c|}{\cellcolor[HTML]{329A9D}{\color[HTML]{000000} \textit{srsRAN}}}                                                                               & \multicolumn{2}{c|}{\cellcolor[HTML]{329A9D}{\color[HTML]{000000} \textit{OAI}}}                                     \\ \cline{2-7} 
\rowcolor[HTML]{68CBD0} 
\multicolumn{1}{|c|}{\multirow{-2}{*}{\cellcolor[HTML]{329A9D}Test \#}}                    & \multicolumn{1}{c|}{\cellcolor[HTML]{68CBD0}{\color[HTML]{000000} \textit{Location}}} & \multicolumn{1}{c|}{\cellcolor[HTML]{68CBD0}{\color[HTML]{000000} \textit{UE Type}}} & \multicolumn{1}{c|}{\cellcolor[HTML]{68CBD0}{\color[HTML]{000000} \textit{DL}}} & \multicolumn{1}{c|}{\cellcolor[HTML]{68CBD0}{\color[HTML]{000000} \textit{UL}}} & \multicolumn{1}{c|}{\cellcolor[HTML]{68CBD0}{\color[HTML]{000000} \textit{DL}}} & {\color[HTML]{000000} \textit{UL}} \\ \hline \hline
\rowcolor[HTML]{ECF4FF} 
\multicolumn{7}{|c|}{\cellcolor[HTML]{ECF4FF}\textbf{Single-UE scenarios}}                                                                                                                                                                                                                                                                                                                                                                                                                                                                                           \\ \hline \hline
\multicolumn{1}{|c|}{\cellcolor[HTML]{99CCFF}T1}                                           & \multicolumn{1}{c|}{\cellcolor[HTML]{99CCFF}{\color[HTML]{000000} A1}}                & \multicolumn{1}{c|}{\cellcolor[HTML]{99CCFF}{\color[HTML]{000000} UE1}}              & \multicolumn{1}{c|}{{\color[HTML]{000000} 86}}                                  & \multicolumn{1}{c|}{{\color[HTML]{000000} 39.1}}                                & \multicolumn{1}{c|}{{\color[HTML]{000000} 101}}                                 & {\color[HTML]{000000} 24.5}        \\ \hline \hline
\multicolumn{1}{|c|}{\cellcolor[HTML]{99CCFF}T2}                                           & \multicolumn{1}{c|}{\cellcolor[HTML]{99CCFF}{\color[HTML]{000000} A2}}                & \multicolumn{1}{c|}{\cellcolor[HTML]{99CCFF}{\color[HTML]{000000} UE2}}              & \multicolumn{1}{c|}{{\color[HTML]{000000} 92.1}}                                & \multicolumn{1}{c|}{{\color[HTML]{000000} 39.9}}                                & \multicolumn{1}{c|}{{\color[HTML]{000000} 109}}                                 & {\color[HTML]{000000} 24.4}        \\ \hline \hline
\multicolumn{1}{|c|}{\cellcolor[HTML]{99CCFF}T3}                                           & \multicolumn{1}{c|}{\cellcolor[HTML]{99CCFF}{\color[HTML]{000000} A3}}                & \multicolumn{1}{c|}{\cellcolor[HTML]{99CCFF}{\color[HTML]{000000} UE3}}              & \multicolumn{1}{c|}{{\color[HTML]{000000} 92.3}}                                & \multicolumn{1}{c|}{{\color[HTML]{000000} 41.3}}                                & \multicolumn{1}{c|}{{\color[HTML]{000000} 102}}                                 & {\color[HTML]{000000} 20.3}        \\ \hline \hline
\multicolumn{1}{|c|}{\cellcolor[HTML]{005AC8}{\color[HTML]{FFFFFF}T4}}                                           & \multicolumn{1}{c|}{\cellcolor[HTML]{005AC8}{\color[HTML]{000000} {\color[HTML]{FFFFFF}D}}}                 & \multicolumn{1}{c|}{\cellcolor[HTML]{005AC8}{\color[HTML]{000000} {\color[HTML]{FFFFFF}UE1}}}              & \multicolumn{1}{c|}{{\color[HTML]{000000} 7.66}}                                & \multicolumn{1}{c|}{{\color[HTML]{000000} 3.23}}                                & \multicolumn{1}{c|}{{\color[HTML]{000000} 48}}                                  & {\color[HTML]{000000} 15.9}        \\ \hline \hline
\multicolumn{1}{|c|}{\cellcolor[HTML]{005AC8}{\color[HTML]{FFFFFF}T5}}                                           & \multicolumn{1}{c|}{\cellcolor[HTML]{005AC8}{\color[HTML]{000000} {\color[HTML]{FFFFFF}D}}}                 & \multicolumn{1}{c|}{\cellcolor[HTML]{005AC8}{\color[HTML]{000000} {\color[HTML]{FFFFFF}UE3}}}              & \multicolumn{1}{c|}{{\color[HTML]{000000} 51.9}}                                & \multicolumn{1}{c|}{{\color[HTML]{000000} 30.3}}                                & \multicolumn{1}{c|}{{\color[HTML]{000000} 66}}                                  & {\color[HTML]{000000} 19.5}        \\ \hline \hline
\multicolumn{1}{|c|}{\cellcolor[HTML]{003161}{\color[HTML]{FFFFFF}T6}}                                           & \multicolumn{1}{c|}{\cellcolor[HTML]{003161}{\color[HTML]{FFFFFF} E}}                 & \multicolumn{1}{c|}{\cellcolor[HTML]{003161}{\color[HTML]{FFFFFF} UE1}}              & \multicolumn{1}{c|}{{\color[HTML]{000000} 3.03}}                                & \multicolumn{1}{c|}{{\color[HTML]{000000} 1.5}}                                 & \multicolumn{1}{c|}{{\color[HTML]{000000} 23.3}}                               & {\color[HTML]{000000} 15.6}       \\ \hline \hline
\multicolumn{1}{|c|}{\cellcolor[HTML]{003161}{\color[HTML]{FFFFFF}T7}}                                           & \multicolumn{1}{c|}{\cellcolor[HTML]{003161}{\color[HTML]{000000} {\color[HTML]{FFFFFF}E}}}                 & \multicolumn{1}{c|}{\cellcolor[HTML]{003161}{\color[HTML]{000000} {\color[HTML]{FFFFFF}UE3}}}              & \multicolumn{1}{c|}{{\color[HTML]{000000} 23.1}}                                & \multicolumn{1}{c|}{{\color[HTML]{000000} 29.3}}                                & \multicolumn{1}{c|}{{\color[HTML]{000000} 35.1}}                                & {\color[HTML]{000000} 14.3}        \\ \hline \hline
\rowcolor[HTML]{ECF4FF} 
\multicolumn{7}{|c|}{\cellcolor[HTML]{ECF4FF}\textbf{Two-UE scenarios}}                                                                                                                                                                                                                                                                                                                                                                                                                                                                                              \\ \hline \hline
\multicolumn{1}{|c|}{\cellcolor[HTML]{99CCFF}{\color[HTML]{000000} }}                      & \multicolumn{1}{c|}{\cellcolor[HTML]{99CCFF}{\color[HTML]{000000} A1}}                & \multicolumn{1}{c|}{\cellcolor[HTML]{99CCFF}{\color[HTML]{000000} UE1}}              & \multicolumn{1}{c|}{{\color[HTML]{000000} 42.5}}                                & \multicolumn{1}{c|}{{\color[HTML]{000000} 18.9}}                                & \multicolumn{1}{c|}{{\color[HTML]{000000} 58.8}}                                & {\color[HTML]{000000} 13.8}        \\ \cline{2-7} 
\multicolumn{1}{|c|}{\multirow{-2}{*}{\cellcolor[HTML]{99CCFF}{\color[HTML]{000000} T8}}}  & \multicolumn{1}{c|}{\cellcolor[HTML]{99CCFF}{\color[HTML]{000000} A2}}                & \multicolumn{1}{c|}{\cellcolor[HTML]{99CCFF}{\color[HTML]{000000} UE2}}              & \multicolumn{1}{c|}{{\color[HTML]{000000} 45.7}}                                & \multicolumn{1}{c|}{{\color[HTML]{000000} 17.9}}                                & \multicolumn{1}{c|}{{\color[HTML]{000000} 65.1}}                                & {\color[HTML]{000000} 11.8}        \\ \hline \hline
\multicolumn{1}{|c|}{\cellcolor[HTML]{99CCFF}{\color[HTML]{000000} }}                      & \multicolumn{1}{c|}{\cellcolor[HTML]{99CCFF}{\color[HTML]{000000} A1}}                & \multicolumn{1}{c|}{\cellcolor[HTML]{99CCFF}{\color[HTML]{000000} UE1}}              & \multicolumn{1}{c|}{{\color[HTML]{000000} 41.2}}                                & \multicolumn{1}{c|}{{\color[HTML]{000000} 17.6}}                                & \multicolumn{1}{c|}{{\color[HTML]{000000} 62.9}}                                & {\color[HTML]{000000} 15.5}        \\ \cline{2-7} 
\multicolumn{1}{|c|}{\multirow{-2}{*}{\cellcolor[HTML]{99CCFF}{\color[HTML]{000000} T9}}}  & \multicolumn{1}{c|}{\cellcolor[HTML]{99CCFF}{\color[HTML]{000000} A3}}                & \multicolumn{1}{c|}{\cellcolor[HTML]{99CCFF}{\color[HTML]{000000} UE3}}              & \multicolumn{1}{c|}{{\color[HTML]{000000} 45.5}}                                & \multicolumn{1}{c|}{{\color[HTML]{000000} 19.9}}                                & \multicolumn{1}{c|}{{\color[HTML]{000000} 66.8}}                                & {\color[HTML]{000000} 14.3}        \\ \hline \hline
\multicolumn{1}{|c|}{\cellcolor[HTML]{99CCFF}{\color[HTML]{000000} }}                      & \multicolumn{1}{c|}{\cellcolor[HTML]{99CCFF}{\color[HTML]{000000} A2}}                & \multicolumn{1}{c|}{\cellcolor[HTML]{99CCFF}{\color[HTML]{000000} UE2}}              & \multicolumn{1}{c|}{{\color[HTML]{000000} 46.9}}                                & \multicolumn{1}{c|}{{\color[HTML]{000000} 16.8}}                                & \multicolumn{1}{c|}{{\color[HTML]{000000} 57.7}}                                & {\color[HTML]{000000} 12.8}        \\ \cline{2-7} 
\multicolumn{1}{|c|}{\multirow{-2}{*}{\cellcolor[HTML]{99CCFF}{\color[HTML]{000000} T10}}} & \multicolumn{1}{c|}{\cellcolor[HTML]{99CCFF}{\color[HTML]{000000} A3}}                & \multicolumn{1}{c|}{\cellcolor[HTML]{99CCFF}{\color[HTML]{000000} UE3}}              & \multicolumn{1}{c|}{{\color[HTML]{000000} 46.8}}                                & \multicolumn{1}{c|}{{\color[HTML]{000000} 21.2}}                                & \multicolumn{1}{c|}{{\color[HTML]{000000} 60.2}}                                & {\color[HTML]{000000} 15}          \\ \hline \hline
\multicolumn{1}{|c|}{\cellcolor[HTML]{005AC8}{\color[HTML]{FFFFFF} }}                      & \multicolumn{1}{c|}{\cellcolor[HTML]{005AC8}{\color[HTML]{FFFFFF} D}}                 & \multicolumn{1}{c|}{\cellcolor[HTML]{005AC8}{\color[HTML]{FFFFFF} UE1}}              & \multicolumn{1}{c|}{{\color[HTML]{000000} 6.51}}                                & \multicolumn{1}{c|}{{\color[HTML]{000000} 5.28}}                                & \multicolumn{1}{c|}{{\color[HTML]{000000} 19.4}}                                & {\color[HTML]{000000} 10.2}        \\ \cline{2-7} 
\multicolumn{1}{|c|}{\multirow{-2}{*}{\cellcolor[HTML]{005AC8}{\color[HTML]{FFFFFF} T11}}} & \multicolumn{1}{c|}{\cellcolor[HTML]{005AC8}{\color[HTML]{FFFFFF} A3}}                & \multicolumn{1}{c|}{\cellcolor[HTML]{005AC8}{\color[HTML]{FFFFFF} UE3}}              & \multicolumn{1}{c|}{{\color[HTML]{000000} 44.5}}                                & \multicolumn{1}{c|}{{\color[HTML]{000000} 19.9}}                                & \multicolumn{1}{c|}{{\color[HTML]{000000} 52.9}}                                & {\color[HTML]{000000} 12.9}        \\ \hline \hline
\multicolumn{1}{|c|}{\cellcolor[HTML]{005AC8}{\color[HTML]{FFFFFF} }}                      & \multicolumn{1}{c|}{\cellcolor[HTML]{005AC8}{\color[HTML]{FFFFFF} A1}}                & \multicolumn{1}{c|}{\cellcolor[HTML]{005AC8}{\color[HTML]{FFFFFF} UE1}}              & \multicolumn{1}{c|}{{\color[HTML]{000000} 31.8}}                                & \multicolumn{1}{c|}{{\color[HTML]{000000} 17.7}}                                & \multicolumn{1}{c|}{{\color[HTML]{000000} 49.6}}                                & {\color[HTML]{000000} 14.2}        \\ \cline{2-7} 
\multicolumn{1}{|c|}{\multirow{-2}{*}{\cellcolor[HTML]{005AC8}{\color[HTML]{FFFFFF} T12}}} & \multicolumn{1}{c|}{\cellcolor[HTML]{005AC8}{\color[HTML]{FFFFFF} D}}                 & \multicolumn{1}{c|}{\cellcolor[HTML]{005AC8}{\color[HTML]{FFFFFF} UE3}}              & \multicolumn{1}{c|}{{\color[HTML]{000000} 27.8}}                                & \multicolumn{1}{c|}{{\color[HTML]{000000} 9.87}}                                & \multicolumn{1}{c|}{{\color[HTML]{000000} 33.9}}                                & {\color[HTML]{000000} 11}          \\ \hline \hline
\multicolumn{1}{|c|}{\cellcolor[HTML]{005AC8}{\color[HTML]{FFFFFF} }}                      & \multicolumn{1}{c|}{\cellcolor[HTML]{005AC8}{\color[HTML]{FFFFFF} D}}                 & \multicolumn{1}{c|}{\cellcolor[HTML]{005AC8}{\color[HTML]{FFFFFF} UE1}}              & \multicolumn{1}{c|}{{\color[HTML]{000000} 2.18}}                                & \multicolumn{1}{c|}{{\color[HTML]{000000} 3.02}}                                & \multicolumn{1}{c|}{{\color[HTML]{000000} 14.3}}                                & {\color[HTML]{000000} 10.1}        \\ \cline{2-7} 
\multicolumn{1}{|c|}{\multirow{-2}{*}{\cellcolor[HTML]{005AC8}{\color[HTML]{FFFFFF} T13}}} & \multicolumn{1}{c|}{\cellcolor[HTML]{005AC8}{\color[HTML]{FFFFFF} A2}}                & \multicolumn{1}{c|}{\cellcolor[HTML]{005AC8}{\color[HTML]{FFFFFF} UE2}}              & \multicolumn{1}{c|}{{\color[HTML]{000000} 34}}                                  & \multicolumn{1}{c|}{{\color[HTML]{000000} 13}}                                  & \multicolumn{1}{c|}{{\color[HTML]{000000} 50.3}}                                & {\color[HTML]{000000} 11}          \\ \hline \hline
\multicolumn{1}{|c|}{\cellcolor[HTML]{005AC8}{\color[HTML]{FFFFFF} }}                      & \multicolumn{1}{c|}{\cellcolor[HTML]{005AC8}{\color[HTML]{FFFFFF} A2}}                & \multicolumn{1}{c|}{\cellcolor[HTML]{005AC8}{\color[HTML]{FFFFFF} UE2}}              & \multicolumn{1}{c|}{{\color[HTML]{000000} 43.7}}                                & \multicolumn{1}{c|}{{\color[HTML]{000000} 16.8}}                                & \multicolumn{1}{c|}{{\color[HTML]{000000} 53.7}}                                & {\color[HTML]{000000} 11.8}        \\ \cline{2-7} 
\multicolumn{1}{|c|}{\multirow{-2}{*}{\cellcolor[HTML]{005AC8}{\color[HTML]{FFFFFF} T14}}} & \multicolumn{1}{c|}{\cellcolor[HTML]{005AC8}{\color[HTML]{FFFFFF} D}}                 & \multicolumn{1}{c|}{\cellcolor[HTML]{005AC8}{\color[HTML]{FFFFFF} UE3}}              & \multicolumn{1}{c|}{{\color[HTML]{000000} 16.3}}                                & \multicolumn{1}{c|}{{\color[HTML]{000000} 14.9}}                                & \multicolumn{1}{c|}{{\color[HTML]{000000} 35.8}}                                & {\color[HTML]{000000} 12.6}        \\ \hline \hline
\rowcolor[HTML]{ECF4FF} 
\multicolumn{7}{|c|}{\cellcolor[HTML]{ECF4FF}\textbf{Three-UE scenarios}}                                                                                                                                                                                                                                                                                                                                                                                                                                                                                            \\ \hline \hline
\multicolumn{1}{|c|}{\cellcolor[HTML]{99CCFF}{\color[HTML]{000000} }}                      & \multicolumn{1}{c|}{\cellcolor[HTML]{99CCFF}{\color[HTML]{000000} A1}}                & \multicolumn{1}{c|}{\cellcolor[HTML]{99CCFF}{\color[HTML]{000000} UE1}}              & \multicolumn{1}{c|}{{\color[HTML]{000000} 26}}                                  & \multicolumn{1}{c|}{{\color[HTML]{000000} 10.2}}                                & \multicolumn{1}{c|}{{\color[HTML]{000000} 41.2}}                                & {\color[HTML]{000000} 11.3}        \\ \cline{2-7} 
\multicolumn{1}{|c|}{\cellcolor[HTML]{99CCFF}{\color[HTML]{000000} }}                      & \multicolumn{1}{c|}{\cellcolor[HTML]{99CCFF}{\color[HTML]{000000} A2}}                & \multicolumn{1}{c|}{\cellcolor[HTML]{99CCFF}{\color[HTML]{000000} UE2}}              & \multicolumn{1}{c|}{{\color[HTML]{000000} 30.6}}                                & \multicolumn{1}{c|}{{\color[HTML]{000000} 10.9}}                                & \multicolumn{1}{c|}{{\color[HTML]{000000} 51.7}}                                & {\color[HTML]{000000} 8.44}        \\ \cline{2-7} 
\multicolumn{1}{|c|}{\multirow{-3}{*}{\cellcolor[HTML]{99CCFF}{\color[HTML]{000000} T15}}} & \multicolumn{1}{c|}{\cellcolor[HTML]{99CCFF}{\color[HTML]{000000} A3}}                & \multicolumn{1}{c|}{\cellcolor[HTML]{99CCFF}{\color[HTML]{000000} UE3}}              & \multicolumn{1}{c|}{{\color[HTML]{000000} 30.9}}                                & \multicolumn{1}{c|}{{\color[HTML]{000000} 14.2}}                                & \multicolumn{1}{c|}{{\color[HTML]{000000} 53.1}}                                & {\color[HTML]{000000} 9.61}        \\ \hline \hline
\multicolumn{1}{|c|}{\cellcolor[HTML]{001C2D}{\color[HTML]{FFFFFF} }}                      & \multicolumn{1}{c|}{\cellcolor[HTML]{001C2D}{\color[HTML]{FFFFFF} D}}                 & \multicolumn{1}{c|}{\cellcolor[HTML]{001C2D}{\color[HTML]{FFFFFF} UE1}}              & \multicolumn{1}{c|}{{\color[HTML]{000000} 5.19}}                                & \multicolumn{1}{c|}{{\color[HTML]{000000} 4.22}}                                & \multicolumn{1}{c|}{{\color[HTML]{000000} 14.3}}                                & {\color[HTML]{000000} 6.5}         \\ \cline{2-7} 
\multicolumn{1}{|c|}{\cellcolor[HTML]{001C2D}{\color[HTML]{FFFFFF} }}                      & \multicolumn{1}{c|}{\cellcolor[HTML]{001C2D}{\color[HTML]{FFFFFF} A2}}                & \multicolumn{1}{c|}{\cellcolor[HTML]{001C2D}{\color[HTML]{FFFFFF} UE2}}              & \multicolumn{1}{c|}{{\color[HTML]{000000} 20.1}}                                & \multicolumn{1}{c|}{{\color[HTML]{000000} 12.6}}                                & \multicolumn{1}{c|}{{\color[HTML]{000000} 38.9}}                                & {\color[HTML]{000000} 5.88}        \\ \cline{2-7} 
\multicolumn{1}{|c|}{\multirow{-3}{*}{\cellcolor[HTML]{001C2D}{\color[HTML]{FFFFFF} T16}}} & \multicolumn{1}{c|}{\cellcolor[HTML]{001C2D}{\color[HTML]{FFFFFF} E}}                 & \multicolumn{1}{c|}{\cellcolor[HTML]{001C2D}{\color[HTML]{FFFFFF} UE3}}              & \multicolumn{1}{c|}{{\color[HTML]{000000} 8.16}}                                & \multicolumn{1}{c|}{{\color[HTML]{000000} 7.22}}                                & \multicolumn{1}{c|}{{\color[HTML]{000000} 13.3}}                                & {\color[HTML]{000000} 7.79}        \\ \hline \hline
\multicolumn{1}{|c|}{\cellcolor[HTML]{001C2D}{\color[HTML]{FFFFFF} }}                      & \multicolumn{1}{c|}{\cellcolor[HTML]{001C2D}{\color[HTML]{FFFFFF} E}}                 & \multicolumn{1}{c|}{\cellcolor[HTML]{001C2D}{\color[HTML]{FFFFFF} UE1}}              & \multicolumn{1}{c|}{{\color[HTML]{000000} X}}                                   & \multicolumn{1}{c|}{{\color[HTML]{000000} X}}                                   & \multicolumn{1}{c|}{{\color[HTML]{000000} X}}                                   & {\color[HTML]{000000} X}           \\ \cline{2-7} 
\multicolumn{1}{|c|}{\cellcolor[HTML]{001C2D}{\color[HTML]{FFFFFF} }}                      & \multicolumn{1}{c|}{\cellcolor[HTML]{001C2D}{\color[HTML]{FFFFFF} A2}}                & \multicolumn{1}{c|}{\cellcolor[HTML]{001C2D}{\color[HTML]{FFFFFF} UE2}}              & \multicolumn{1}{c|}{{\color[HTML]{000000} X}}                                   & \multicolumn{1}{c|}{{\color[HTML]{000000} X}}                                   & \multicolumn{1}{c|}{{\color[HTML]{000000} X}}                                   & {\color[HTML]{000000} X}           \\ \cline{2-7} 
\multicolumn{1}{|c|}{\multirow{-3}{*}{\cellcolor[HTML]{001C2D}{\color[HTML]{FFFFFF} T17}}} & \multicolumn{1}{c|}{\cellcolor[HTML]{001C2D}{\color[HTML]{FFFFFF} D}}                 & \multicolumn{1}{c|}{\cellcolor[HTML]{001C2D}{\color[HTML]{FFFFFF} UE3}}              & \multicolumn{1}{c|}{{\color[HTML]{000000} X}}                                   & \multicolumn{1}{c|}{{\color[HTML]{000000} X}}                                   & \multicolumn{1}{c|}{{\color[HTML]{000000} X}}                                   & {\color[HTML]{000000} X}           \\ \hline
\end{tabular}%
}
\end{table}


\subsubsection{Results on tests conducted on \textit{``good''} locations}
\begin{itemize}[wide, labelwidth=!, labelindent=0pt]
\item \emph{Impact of the RAN software:} 
        Throughout our tests we observed  that \textit{srsRAN}  delivers much higher \gls{UL} rates, while \textit{OAI-RAN} performs better on the \gls{DL}, regardless of the type of the \gls{SDR} and the number or the type of connected \gls{UE}s.
        \item \emph{Impact of the type of UE:} 
        Our tests indicate that for almost all of the single/multi-\gls{UE} scenarios, irrespective of the \gls{RAN} software, the two modem-based \gls{UE}s (UE$_2$, and UE$_3$), outperform the phone in terms of \gls{DL} rates.
        Focusing on \textit{srsRAN}, we observe that in single-\gls{UE} scenarios (T1 vs. T2 \& T3), UE$_1$ is receiving 93\% of the \gls{DL} rate of UE$_2$ and UE$_3$. Similarly, UE$_1$ receives 92\% and 90\% of the \gls{DL} rate of UE$_2$ and UE$_3$, when two \gls{UE}s are connected at the same time, in T8 \& T9. When it comes to T15 (corresponding to the three UEs case), UE$_1$ is receiving only 84\% of the \gls{DL} rate achieved by the other two modem-based \gls{UE}s. This pattern is also evident  in the results achieved by \textit{OAI-RAN} even if in the single-\gls{UE} scenarios, the difference between the \gls{DL} rates of UE$_1$ and UE$_3$ is negligible. Indeed, there is an 8\% gap between the \gls{DL} rates of UE$_1$ and UE$_2$. Moving to the multi-\gls{UE} scenarios with \textit{OAI-RAN}, we see that the gap between the \gls{DL} rate of UE$_1$ and the other two modem-based \gls{UE}s increases. In two-\gls{UE} scenarios, T8 and T9, UE$_1$ achieved 90\% and 94\% of the \gls{DL} rate of UE$_2$ and UE$_3$, respectively. When all \gls{UE}s are connected in T15, UE$_1$ is only able to receive 79\% and 77\% of the \gls{DL} rate achieved by UE$_2$ and UE$_3$, respectively. 
        The results on the \gls{UL} are difficult to interpret since on single-UE scenarios, UE$_3$ does better than the other UEs for \textit{srsRAN} and worse for \textit{OAI-RAN}.

        \item \emph{Cases with multiple UEs:} Throughout our multi-\gls{UE} tests in \textit{``good''} locations, we observed that both $S_{RAN}$ do seem to share the resources \emph{roughly} equally among the \gls{UE}s in both \gls{UL}, and \gls{DL} directions. For instance, comparing T3, T8, and T15 in the \gls{UL} direction, we see that if \textit{srsRAN} is used, the maximum  \gls{UL} rate achieved in a single-\gls{UE} scenario is 41.3 Mbps. In the two-\gls{UE} scenario (T8), UE$_1$ and UE$_2$  receive 45\% and 43\% of this rate, respectively. Moreover, in T15, when all \gls{UE}s are connected, UE$_1$, UE$_2$, and UE$_3$, they can send 24\%, 26\%, and 34\% of the maximum achieved \gls{UL} rate in the single-\gls{UE} scenario. If \textit{OAI-RAN} is used, the maximum  \gls{UL} rate achieved in single-\gls{UE} scenarios is 24.5 Mbps. We see that in T8, UE$_1$ and UE$_2$ each can send 56\% and 48\% of this maximum rate, respectively. In T15, UE$_1$, UE$_2$, and UE$_3$ each can transmit 46\%, 34\%, and 39\% of the maximum \gls{UL} rate. While \textit{OAI-RAN} performs poorly in the \gls{UL}, it shows better resource sharing capabilities in multi-\gls{UE} scenarios.
    \end{itemize}

\subsubsection{Observations on tests conducted on \textit{``good''} locations}
\begin{itemize}[wide, labelwidth=!, labelindent=0pt]
        \item \emph{Impact of the type of UE:} 
        Overall, we found it easier to work with modem-based \gls{UE}s. While \textit{srsRAN} did not exhibit any apparent differences in the attachment process of the \gls{UE}s, when working with \textit{OAI-RAN}, we observed that the phone, UE$_1$, had a harder time attaching to the testbed and maintaining its connection for three minutes during the tests. We did not observe such a behaviour with the two modem-based \gls{UE}s, while connected to \textit{OAI-RAN}.

        \item \emph{Effect of multiple UEs:} Throughout our multi-\gls{UE} tests in \textit{``good''} locations, we noticed that \textit{OAI-RAN} crashed several times, 
        specifically when the tests were done on the \gls{UL} direction.

    \end{itemize}

\subsubsection{Results on tests conducted on \textit{``fair''} and \textit{``bad''} locations}
    \begin{itemize}[wide, labelwidth=!, labelindent=0pt]
        \item \emph{Impact of location}: The overall observation is that 
        there is a significant difference between the two \gls{RAN} software platforms: while OAI-RAN adjusts the power automatically, srsRAN provides a static power setting mechanism, which is through setting the Tx/Rx gain values. 
        This is difficult to adjust in scenarios with multiple \gls{UE}s.
        We carefully selected the Tx and Rx gains that resulted in the best achieved rates in ``good'' positions, and maintained those values for the gNB throughout all our tests for all testbeds, and in that case, \textit{srsRAN} loses its superior performance, in the presence of multi-\gls{UE}s, each located on a different type of position.
        As for \textit{OAI-RAN}, the automatic power adjustment feature provides some consistency in the results based on the locations of the \gls{UE}s. We see from T1, T4, and T6 that moving UE$_1$ from A1 to D, and to E causes a 52\% and 76\% drop in the \gls{DL} data rate, and 35\%, and 36\% in the \gls{UL} rate, respectively. Looking at UE$_3$ (T3, T5, T7), we see that the drop in the \gls{DL} (\gls{UL} respectively) rates from point A3 to D is 40\% (20\% respectively), and from A3 to E is 67\% (40\% respectively). \textit{OAI-RAN}'s performance is thus not impacted by the location of the connected \gls{UE} in the \gls{UL} direction as much as it is in the \gls{DL}. The same trend is also seen in the multi-\gls{UE} scenarios with \textit{OAI-RAN}.
        
        \item \emph{Impact of the type of UE:} The most unexpected observation for us was the fact that srsRAN seems to provide very limited coverage for UE$_1$. Comparing T4 and T5, as well as comparing T6 and T7, you can see that while UE$_3$ is able to achieve 51 and 23 Mbps in DL, the phone is only getting around 7 and 3 Mbps respectively. 
        This trend is also evident in multi-UE scenarios with srsRAN. On the OAI-RAN side, we did observe this trend, but it was much less pronounced. For instance, comparing T4 and T5, we see that the phone is getting 48 Mbps in the DL, while the modem is achieving around 66 Mbps.
        There is certainly a difference, but it is nowhere near as drastic as the gap observed with srsRAN.

        \item \emph{Effect of multiple UEs:} 
        As mentioned in the previous points, since srsRAN does not support automatic power adjustment, the results in \textit{``fair''} and \textit{``bad''} locations are very poor. Additionally, the unforeseeable discrepancy between the performance seen by UE$_1$ and UE$_3$ in those locations has made this category of results for srsRAN to be haphazard and inconsistent, not revealing any kind of pattern. However, \textit{OAI-RAN} was more reliable. \textit{OAI-RAN} seems to share the Physical Resource Blocks (PRBs) equally among users, and the final rate achieved by the UE is then determined by the \gls{MCS} value. For instance, taking a look at T3, T4, and T11, we see that in the case of T11, the two-UE scenario, each \gls{UE} is receiving roughly half of what they used to get at the same location in the single-\gls{UE} scenario, e.g., UE$_1$ which was receiving about 48 Mbps at D in the single-\gls{UE} scenario, is now receiving 19.2 Mbps when UE$_3$ is also connected from location A3, and UE$_3$ which was receiving about 102 Mbps at A3 in the sinlge-\gls{UE} scenario, is now receiving 52.9 Mbps at location D, in the two-\gls{UE} scenario.
    \end{itemize}
    
\subsubsection{Observations on tests conducted on \textit{``fair''} and \textit{``bad''} locations}

\begin{itemize}[wide, labelwidth=!, labelindent=0pt]
        \item \emph{Effect of multiple UEs:} 
        We have to mention that for both software platforms, we were unable to conduct T17. Despite attempting the test more than seven times, UE$_1$ was unable to maintain its connection for the duration of the test (three minutes) for each data rate test at location $E$.
\end{itemize}

\subsubsection{Second round of tests to compare the SDRs}
For the second round of tests, we configured all the tests to be done on \textit{``good''} locations.
The results are presented in Table~\ref{tab:SDR}.
In our tests, we observed that between the two \gls{SDR}s, \textit{\gls{USRP} X410} yielded better \gls{UL} rates. Based on the results presented, \textit{OAI-RAN} shows the largest improvement in \gls{UL} rates when \textit{USRP X410} is utilized (up to 48\% improvement). For instance, in a single-\gls{UE} scenario with UE$_3$, using \textit{\{USRP B210, OAI-RAN\}} results in 20.3 Mbps in \gls{UL}, whereas using \textit{\{USRP X410, OAI-RAN\}} results in a 45\% improvement, reaching 29.5 Mbps. In the same scenario, if \textit{srsRAN} is used, changing the \gls{SDR} from \textit{USRP B210} to \textit{USRP X410} results in no improvement.
Additionally, during single-\gls{UE} scenarios, we observe the best \gls{DL} performance with \textit{USRP X410}.
As an example, when UE$_1$ is the only connected \gls{UE}, the combination of \textit{\{USRP X410, OAI-RAN\}} results in 13\% improvement in \gls{DL} rate, compared to \textit{\{USRP B210, OAI-RAN\}}. For the same scenario and \gls{UE}, the combination of \textit{\{USRP X410, srsRAN\}} improves the \gls{DL} rate by 8\% compared to \textit{\{USRP B210, srsRAN\}}.



\begin{table}[t]
\caption{Data rate results for two SDRs (in Mbps)}
\label{tab:SDR}
\resizebox{\columnwidth}{!}{%
\begin{tabular}{|ccccccccc|}
\hline
\rowcolor[HTML]{68CBD0} 
\multicolumn{1}{|c|}{\cellcolor[HTML]{68CBD0}{\color[HTML]{000000} \textit{SDR}}}    & \multicolumn{4}{c|}{\cellcolor[HTML]{68CBD0}{\color[HTML]{000000} \textit{USRP B210}}}                                                                                                                                                                                                                                                & \multicolumn{4}{c|}{\cellcolor[HTML]{68CBD0}{\color[HTML]{000000} \textit{USRP X410}}}                                                                                                                                                                                                   \\ \hline
\rowcolor[HTML]{68CBD0} 
\multicolumn{1}{|c|}{\cellcolor[HTML]{68CBD0}{\color[HTML]{000000} \textbf{$S_{RAN}$}}} & \multicolumn{2}{c|}{\cellcolor[HTML]{68CBD0}{\color[HTML]{000000} \textit{srsRAN}}}                                                                               & \multicolumn{2}{c|}{\cellcolor[HTML]{68CBD0}{\color[HTML]{000000} \textit{OAI}}}                                                                                  & \multicolumn{2}{c|}{\cellcolor[HTML]{68CBD0}{\color[HTML]{000000} \textit{srsRAN}}}                                                                               & \multicolumn{2}{c|}{\cellcolor[HTML]{68CBD0}{\color[HTML]{000000} \textit{OAI}}}                                     \\ \hline
\rowcolor[HTML]{329A9D} 
\multicolumn{1}{|c|}{\cellcolor[HTML]{329A9D}{\color[HTML]{000000} }}                & \multicolumn{1}{c|}{\cellcolor[HTML]{329A9D}{\color[HTML]{000000} \textit{DL}}} & \multicolumn{1}{c|}{\cellcolor[HTML]{329A9D}{\color[HTML]{000000} \textit{UL}}} & \multicolumn{1}{c|}{\cellcolor[HTML]{329A9D}{\color[HTML]{000000} \textit{DL}}} & \multicolumn{1}{c|}{\cellcolor[HTML]{329A9D}{\color[HTML]{000000} \textit{UL}}} & \multicolumn{1}{c|}{\cellcolor[HTML]{329A9D}{\color[HTML]{000000} \textit{DL}}} & \multicolumn{1}{c|}{\cellcolor[HTML]{329A9D}{\color[HTML]{000000} \textit{UL}}} & \multicolumn{1}{c|}{\cellcolor[HTML]{329A9D}{\color[HTML]{000000} \textit{DL}}} & {\color[HTML]{000000} \textit{UL}} \\ \hline
\rowcolor[HTML]{ECF4FF} 
\multicolumn{9}{|c|}{\cellcolor[HTML]{ECF4FF}\textbf{Single-UE scenarios}}                                                                                                                                                                                                                                                                                                                                                                                                                                                                                                                                                                                                                             \\ \hline \hline
\multicolumn{1}{|c|}{\cellcolor[HTML]{99CCFF}{\color[HTML]{000000} UE1}}             & \multicolumn{1}{c|}{86}                                                         & \multicolumn{1}{c|}{39.1}                                                       & \multicolumn{1}{c|}{101}                                                        & \multicolumn{1}{c|}{24.5}                                                       & \multicolumn{1}{c|}{92.9}                                                       & \multicolumn{1}{c|}{40.4}                                                       & \multicolumn{1}{c|}{115}                                                        & 25.5                               \\ \hline \hline
\multicolumn{1}{|c|}{\cellcolor[HTML]{339FFF}{\color[HTML]{000000} UE2}}             & \multicolumn{1}{c|}{92.1}                                                       & \multicolumn{1}{c|}{39.9}                                                       & \multicolumn{1}{c|}{109}                                                        & \multicolumn{1}{c|}{24.4}                                                       & \multicolumn{1}{c|}{92.1}                                                       & \multicolumn{1}{c|}{40.9}                                                       & \multicolumn{1}{c|}{115}                                                        & 33.3                               \\ \hline \hline
\multicolumn{1}{|c|}{\cellcolor[HTML]{0074FF}{\color[HTML]{000000} UE3}}             & \multicolumn{1}{c|}{92.3}                                                       & \multicolumn{1}{c|}{41.3}                                                       & \multicolumn{1}{c|}{102}                                                        & \multicolumn{1}{c|}{20.3}                                                       & \multicolumn{1}{c|}{92.9}                                                       & \multicolumn{1}{c|}{41.3}                                                       & \multicolumn{1}{c|}{115}                                                        & 29.5                               \\ \hline \hline
\rowcolor[HTML]{ECF4FF} 
\multicolumn{9}{|c|}{\cellcolor[HTML]{ECF4FF}{\color[HTML]{000000} \textbf{Two-UE scenarios}}}                                                                                                                                                                                                                                                                                                                                                                                                                                                                                                                                                                                                      \\ \hline \hline
\multicolumn{1}{|c|}{\cellcolor[HTML]{005AC8}{\color[HTML]{FFFFFF} UE1}}             & \multicolumn{1}{c|}{42.5}                                                       & \multicolumn{1}{c|}{18.9}                                                       & \multicolumn{1}{c|}{58.8}                                                       & \multicolumn{1}{c|}{13.8}                                                       & \multicolumn{1}{c|}{45.3}                                                       & \multicolumn{1}{c|}{17.7}                                                       & \multicolumn{1}{c|}{66.9}                                                       & 17.2                               \\ \hline
\multicolumn{1}{|c|}{\cellcolor[HTML]{005AC8}{\color[HTML]{FFFFFF} UE2}}             & \multicolumn{1}{c|}{45.7}                                                       & \multicolumn{1}{c|}{17.9}                                                       & \multicolumn{1}{c|}{65.1}                                                       & \multicolumn{1}{c|}{11.8}                                                       & \multicolumn{1}{c|}{46.8}                                                       & \multicolumn{1}{c|}{{\color[HTML]{000000} 21.3}}                                & \multicolumn{1}{c|}{66.9}                                                       & 17.5                               \\ \hline \hline
\multicolumn{1}{|c|}{\cellcolor[HTML]{004594}{\color[HTML]{FFFFFF} UE1}}             & \multicolumn{1}{c|}{41.2}                                                       & \multicolumn{1}{c|}{17.6}                                                       & \multicolumn{1}{c|}{62.9}                                                       & \multicolumn{1}{c|}{15.5}                                                       & \multicolumn{1}{c|}{48}                                                         & \multicolumn{1}{c|}{16.7}                                                       & \multicolumn{1}{c|}{62.2}                                                       & 18.2                               \\ \hline
\multicolumn{1}{|c|}{\cellcolor[HTML]{004594}{\color[HTML]{FFFFFF} UE3}}             & \multicolumn{1}{c|}{45.5}                                                       & \multicolumn{1}{c|}{19.9}                                                       & \multicolumn{1}{c|}{66.8}                                                       & \multicolumn{1}{c|}{14.3}                                                       & \multicolumn{1}{c|}{44.5}                                                       & \multicolumn{1}{c|}{20.2}                                                       & \multicolumn{1}{c|}{62.6}                                                       & 16.5                               \\ \hline \hline
\multicolumn{1}{|c|}{\cellcolor[HTML]{003161}{\color[HTML]{FFFFFF} UE2}}             & \multicolumn{1}{c|}{46.9}                                                       & \multicolumn{1}{c|}{16.8}                                                       & \multicolumn{1}{c|}{57.7}                                                       & \multicolumn{1}{c|}{12.8}                                                       & \multicolumn{1}{c|}{48.5}                                                       & \multicolumn{1}{c|}{21.3}                                                       & \multicolumn{1}{c|}{61.3}                                                       & 16.4                               \\ \hline
\multicolumn{1}{|c|}{\cellcolor[HTML]{003161}{\color[HTML]{FFFFFF} UE3}}             & \multicolumn{1}{c|}{46.8}                                                       & \multicolumn{1}{c|}{21.2}                                                       & \multicolumn{1}{c|}{60.2}                                                       & \multicolumn{1}{c|}{15}                                                         & \multicolumn{1}{c|}{45.1}                                                       & \multicolumn{1}{c|}{20}                                                         & \multicolumn{1}{c|}{61.4}                                                       & 14.9                               \\ \hline \hline
\rowcolor[HTML]{ECF4FF} 
\multicolumn{9}{|c|}{\cellcolor[HTML]{ECF4FF}{\color[HTML]{000000} \textbf{Three-UE scenario}}}                                                                                                                                                                                                                                                                                                                                                                                                                                                                                                                                                                                                      \\  \hline \hline
\multicolumn{1}{|c|}{\cellcolor[HTML]{001C2D}{\color[HTML]{FFFFFF} UE1}}             & \multicolumn{1}{c|}{26}                                                         & \multicolumn{1}{c|}{10.2}                                                       & \multicolumn{1}{c|}{41.2}                                                       & \multicolumn{1}{c|}{11.3}                                                       & \multicolumn{1}{c|}{28.4}                                                       & \multicolumn{1}{c|}{13}                                                         & \multicolumn{1}{c|}{46.1}                                                       & 8.1       \\ \hline
\multicolumn{1}{|c|}{\cellcolor[HTML]{001C2D}{\color[HTML]{FFFFFF} UE2}}             & \multicolumn{1}{c|}{30.6}                                                       & \multicolumn{1}{c|}{10.9}                                                       & \multicolumn{1}{c|}{51.7}                                                       & \multicolumn{1}{c|}{8.44}                                                       & \multicolumn{1}{c|}{24.3}                                                       & \multicolumn{1}{c|}{13.8}                                                       & \multicolumn{1}{c|}{46.1}                                                       & 11.3       \\ \hline
\multicolumn{1}{|c|}{\cellcolor[HTML]{001C2D}{\color[HTML]{FFFFFF} UE3}}             & \multicolumn{1}{c|}{30.9}                                                       & \multicolumn{1}{c|}{14.2}                                                       & \multicolumn{1}{c|}{53.1}                                                       & \multicolumn{1}{c|}{9.61}                                                       & \multicolumn{1}{c|}{30.8}                                                       & \multicolumn{1}{c|}{13.9}                                                       & \multicolumn{1}{c|}{46}                                                         & 10.3       \\ \hline
\end{tabular}%
}
\end{table}
\subsection{Methodology \& Results for Latency Assessment}
\subsubsection{Methodology}
To analyze the \gls{E2E} latency, we followed the same methodology as for the data rate assessments, as in we used the same testbeds, elements and configurations described in the data rate assessments. We also used the same single/multi-\gls{UE} scenarios with different types of locations.
The latency tests were done using the \textit{ping} command.
For each assessment, a \textit{ping} command from the measuring \gls{UE} to the IP address of the AMF in the core network was conducted.
Based on our observations, having multiple connected \gls{UE}s did not affect the \gls{E2E} latency experienced by each \gls{UE}. Furthermore, the positions of the connected \gls{UE}s also did not appear to influence the final result. Therefore, to simplify comparisons, we only report on the latencies achieved by different types of UEs in single-UE scenarios, where each UE is located in a \textit{``good''} position. Table~\ref{tab:Latency} presents our results.
    

\subsubsection{Impact of the RAN software}
Similar to the data rate tests, we observe the impact of the choice of $S_{RAN}$ on the \gls{E2E} latency. Although recent updates for both software platforms have brought the data rates closer together, a significant difference in the latency results persists. \textit{OAI-RAN} outperforms \textit{srsRAN} in this regard.
Comparing the latency results achieved by UE$_1$, when connected to \textit{USRP B210}, we find that when we have the shortest \gls{E2E} latency results for both $S_{RAN}$, the result achieved by \textit{OAI-RAN} is 70\% shorter than that of \textit{srsRAN}.
\subsubsection{Impact of the type of UE} 
Table~\ref{tab:Latency} shows that the type of \gls{UE} plays some role in the final \gls{E2E} latency. Notably, UE$_1$ consistently achieved lower latency compared to the modem-based \gls{UE}s. Additionally, within the modem-based \gls{UE}s, UE$_2$, which is a modem connected to a Linux system shows slightly superior performance compared to UE$_3$, the same modem, connected to a Windows laptop.
        
\subsubsection{Comparison between the SDRs} In our tests, we found that other than for two scenarios, the measured latencies were relatively consistent regardless of the \gls{SDR} device used. The two exceptions are, the latency result achieved by UE$_1$ with \textit{\{USRP B210, srsRAN\}} is 6.9\% less than with \textit{\{USRP X410, srsRAN\}}, the latency result achieved by UE$_3$ with \textit{\{USRP B210, OAI-RAN\}} is 13\% less than with \textit{\{USRP X410, OAI-RAN\}}.

\begin{table}[]
\centering
\caption{E2E latency between the UE and 5GC (in ms)}
\label{tab:Latency}
\resizebox{0.85\columnwidth}{!}{%
\begin{tabular}{|c|cc|cc|}
\hline
\rowcolor[HTML]{68CBD0} 
{\color[HTML]{000000} SDR}                & \multicolumn{2}{c|}{\cellcolor[HTML]{68CBD0}{\color[HTML]{000000} \textit{USRP B210}}}                                    & \multicolumn{2}{c|}{\cellcolor[HTML]{68CBD0}{\color[HTML]{000000} \textit{USRP X410}}}                                    \\ \hline
\rowcolor[HTML]{68CBD0} 
{\color[HTML]{000000} $S_{RAN}$}             & \multicolumn{1}{c|}{\cellcolor[HTML]{68CBD0}{\color[HTML]{000000} \textit{srsRAN}}} & {\color[HTML]{000000} \textit{OAI-RAN}} & \multicolumn{1}{c|}{\cellcolor[HTML]{68CBD0}{\color[HTML]{000000} \textit{srsRAN}}} & {\color[HTML]{000000} \textit{OAI-RAN}} \\ \hline \hline
\cellcolor[HTML]{99CCFF}{\color[HTML]{000000} UE1} & \multicolumn{1}{c|}{27}                                                             & 8                                   & \multicolumn{1}{c|}{29}                                                             & 8                                   \\ \hline \hline
\cellcolor[HTML]{339FFF}{\color[HTML]{000000} UE2} & \multicolumn{1}{c|}{31.8}                                                           & 10.4                                & \multicolumn{1}{c|}{31.2}                                                           & 11.3                                \\ \hline \hline
\cellcolor[HTML]{0074FF}{\color[HTML]{000000} UE3} & \multicolumn{1}{c|}{33}                                                             & 13                                  & \multicolumn{1}{c|}{34}                                                             & 15                                  \\ \hline
\end{tabular}%
}
\end{table}

\subsection{Methodology \& Results for Coverage Assessment}
\subsubsection{Methodology}
    
We assess the coverage of four testbeds using UE$_1$, specifically focusing on the impact of the \gls{SDR} and $S_{RAN}$. These included combinations of \{\textit{srsRAN}, \textit{OAI-RAN}\} for the $S_{RAN}$, alongside \{\gls{USRP} B210, \gls{USRP} X410\}. 
All the measurement locations are indicated on the map, as shown in Fig.~\ref{fig:Coverage}, by purple circles. At each location, we utilized the \textit{5G Switch - Force 5G Only} application on UE$_1$ and recorded the \gls{RSRP} values in Table~\ref{tab:Coverage} (Please note that any cell with X as their value in the table, indicate a disconnection).

\subsubsection{Impact of the software} 
Due to the automatic power adjustment feature, \textit{OAI-RAN} outperforms \textit{srsRAN} in terms if coverage support. 
A notable observation was that in areas with very poor signal strength, such as points $L$, $G$, or $M$, \textit{OAI-RAN} exhibited a tendency to crash frequently. In contrast, \textit{srsRAN} gNB at such points continue to run, but the UE is unable to find any connections.
        
\subsubsection{Comparison between the SDRs} Comparing the coverage maps of the two \gls{SDR} devices, it is evident that \textit{USRP X410} exhibits a superior performance. However, the impact of the $S_{RAN}$ is more influential. Specifically, the testbed utilizing \{\textit{OAI-RAN}, \textit{USRP B210}\} exhibits a better efficacy compared to \{\textit{srsRAN}, \textit{USRP X410}\}.


\begin{table}[]
\centering
\caption{Signal strength measurements (in dBm)}
\label{tab:Coverage}
\resizebox{0.7\columnwidth}{!}{%
\begin{tabular}{|
>{\columncolor[HTML]{99CCFF}}c |
>{\columncolor[HTML]{FFFFFF}}c 
>{\columncolor[HTML]{FFFFFF}}c |
>{\columncolor[HTML]{FFFFFF}}c 
>{\columncolor[HTML]{FFFFFF}}c |}
\hline
\cellcolor[HTML]{329A9D}{\color[HTML]{330001} $S_{RAN}$} & \multicolumn{2}{c|}{\cellcolor[HTML]{329A9D}{\color[HTML]{000000} \textit{srsRAN}}}                                                     & \multicolumn{2}{c|}{\cellcolor[HTML]{329A9D}{\color[HTML]{000000} \textit{OAI-RAN}}}                                                    \\ \hline
\cellcolor[HTML]{68CBD0}{\color[HTML]{330001} \textit{SDR}}        & \multicolumn{1}{c|}{\cellcolor[HTML]{68CBD0}{\color[HTML]{000000} \textit{B210}}} & \cellcolor[HTML]{68CBD0}{\color[HTML]{000000} \textit{X410}} & \multicolumn{1}{c|}{\cellcolor[HTML]{68CBD0}{\color[HTML]{000000} \textit{B210}}} & \cellcolor[HTML]{68CBD0}{\color[HTML]{000000} \textit{X410}} \\ \hline \hline
{\color[HTML]{000000} A2}                                 & \multicolumn{1}{c|}{\cellcolor[HTML]{FFFFFF}{\color[HTML]{000000} -100}} & -84                                                 & \multicolumn{1}{c|}{\cellcolor[HTML]{FFFFFF}-92}                         & -74                                                 \\ \hline \hline
{\color[HTML]{000000} B}                                  & \multicolumn{1}{c|}{\cellcolor[HTML]{FFFFFF}-109}                        & -100                                                & \multicolumn{1}{c|}{\cellcolor[HTML]{FFFFFF}-107}                        & -85                                                 \\ \hline \hline
{\color[HTML]{000000} C}                                  & \multicolumn{1}{c|}{\cellcolor[HTML]{FFFFFF}{\color[HTML]{000000} -109}} & -101                                                & \multicolumn{1}{c|}{\cellcolor[HTML]{FFFFFF}-107}                        & -91                                                 \\ \hline \hline
{\color[HTML]{000000} D}                                  & \multicolumn{1}{c|}{\cellcolor[HTML]{FFFFFF}-119}                        & -110                                                & \multicolumn{1}{c|}{\cellcolor[HTML]{FFFFFF}-113}                        & -102                                                \\ \hline \hline
{\color[HTML]{000000} E}                                  & \multicolumn{1}{c|}{\cellcolor[HTML]{FFFFFF}-125}                        & -114                                                & \multicolumn{1}{c|}{\cellcolor[HTML]{FFFFFF}-119}                        & -106                                                \\ \hline \hline
{\color[HTML]{000000} F}                                  & \multicolumn{1}{c|}{\cellcolor[HTML]{FFFFFF}-131}                        & -121                                                & \multicolumn{1}{c|}{\cellcolor[HTML]{FFFFFF}-122}                        & -106                                                \\ \hline \hline
{\color[HTML]{000000} G}                                  & \multicolumn{1}{c|}{\cellcolor[HTML]{FFFFFF}X}                           & -127                                                & \multicolumn{1}{c|}{\cellcolor[HTML]{FFFFFF}X}                           & -108                                                \\ \hline \hline
{\color[HTML]{000000} H}                                  & \multicolumn{1}{c|}{\cellcolor[HTML]{FFFFFF}-124}                        & -115                                                & \multicolumn{1}{c|}{\cellcolor[HTML]{FFFFFF}-116}                        & -106                                                \\ \hline \hline
{\color[HTML]{000000} I}                                  & \multicolumn{1}{c|}{\cellcolor[HTML]{FFFFFF}-130}                        & -119                                                & \multicolumn{1}{c|}{\cellcolor[HTML]{FFFFFF}-120}                        & -111                                                \\ \hline \hline
{\color[HTML]{000000} J}                                  & \multicolumn{1}{c|}{\cellcolor[HTML]{FFFFFF}-118}                        & -113                                                & \multicolumn{1}{c|}{\cellcolor[HTML]{FFFFFF}-115}                        & -99                                                 \\ \hline \hline
{\color[HTML]{000000} K}                                  & \multicolumn{1}{c|}{\cellcolor[HTML]{FFFFFF}-127}                        & -120                                                & \multicolumn{1}{c|}{\cellcolor[HTML]{FFFFFF}-125}                        & -111                                                \\ \hline \hline
{\color[HTML]{000000} L}                                  & \multicolumn{1}{c|}{\cellcolor[HTML]{FFFFFF}-135}                        & -127                                                & \multicolumn{1}{c|}{\cellcolor[HTML]{FFFFFF}X}                           & -114                                                \\ \hline \hline
{\color[HTML]{000000} M}                                  & \multicolumn{1}{c|}{\cellcolor[HTML]{FFFFFF}X}                           & X                                                   & \multicolumn{1}{c|}{\cellcolor[HTML]{FFFFFF}X}                           & -117                                                \\ \hline \hline
{\color[HTML]{000000} N}                                  & \multicolumn{1}{c|}{\cellcolor[HTML]{FFFFFF}X}                           & X                                                   & \multicolumn{1}{c|}{\cellcolor[HTML]{FFFFFF}X}                           & -116                                                \\ \hline \hline
{\color[HTML]{000000} O}                                  & \multicolumn{1}{c|}{\cellcolor[HTML]{FFFFFF}X}                           & X                                                   & \multicolumn{1}{c|}{\cellcolor[HTML]{FFFFFF}X}                           & -120                                                \\ \hline
\end{tabular}%
}
\end{table}

\subsection{Methodology \& Results for Computational Resource Consumption of the Open-source RAN Software }

\subsubsection{Methodology}  
There are two aspects to the analysis of resource consumption for the open-source software platforms. First, how each RAN software consumes the available computational resources, and second, how the choice of the host PC affects the performance achieved by the \gls{UE}s. 
To answer the first question, we set up testbeds using \{\textit{srsRAN}, \textit{OAI-RAN}\} as their $S_{RAN}$, \textit{Open5GS} as $S_{5GC}$, and set the host PC for the software platforms as PC \#1. 
Please note that PC \#1 is our more powerful computer with 16 CPU cores, running at the frequency of 3.5 GHz.
We are specifically interested in determining the maximum resource consumption by each $S_{RAN}$ in a full buffer scenario
in both \gls{UL} and \gls{DL} transmissions, for single and multi-UE scenarios. 
This allows us to gain insight into the worst-case computational consumption scenario for each software platform.

\subsubsection{Observations on CPU utilization}
\begin{figure}[ht]
    \centering
    \includegraphics[width=0.9\columnwidth]{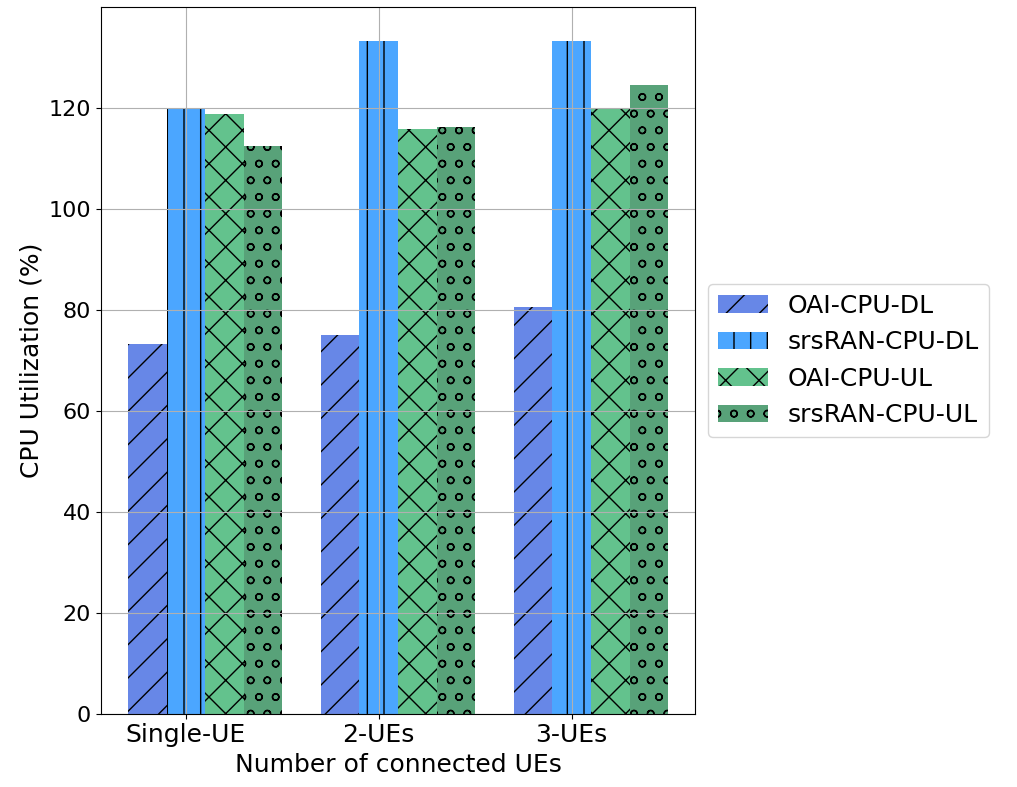}
    \caption{CPU utilization as a function of number of UEs for OAI and srsRAN in UL and DL}
    \label{fig:CRC}
\end{figure}

Please see Fig.~\ref{fig:CRC}, that shows the CPU utilization as a function of the number of connected \gls{UE}s.
\begin{itemize}[wide, labelwidth=!, labelindent=0pt]
    \item \emph{Traffic direction effect, UL vs. DL:}
    At first glance, we observe a trend with \textit{OAI-RAN}: the \gls{DL} traffic appears to be less CPU-hungry than the \gls{UL}. This trend is reversed for \textit{srsRAN}, i.e., the \gls{UL} traffic seems to require fewer CPU cores. Another important observation is the significant difference in computational resource consumption between \textit{srsRAN} and \textit{OAI} in the \gls{DL} direction. With a single UE, OAI consumes 73\% of one CPU core, whereas \textit{srsRAN} consumes 1.2 CPU cores, indicating a drastic gap in resource utilization (Please note that none of the $S_{RAN}$ utilized more than 2 cores of PC \#1).

    \item \emph{Effect of multiple UEs:} There is a gradual increase in the resource consumption for both $S_{RAN}$ as the number of connected \gls{UE}s increases.

\end{itemize}

\subsubsection{Observations on memory utilization} Our results indicate that the CPU is the primary bottleneck resource, while memory usage remains relatively consistent. In all scenarios, \textit{OAI-RAN} consumed a maximum of 3\% of the memory, whereas \textit{srsRAN} never exceeded 10.2\% of the memory.

\subsubsection{Observations on the effect of the host PC} For this round of tests, utilizing the less powerful PC, PC \#2, we established two additional testbeds with the two open-source RAN platforms and UE$_1$. This setup aims to compare the data rate and latency results with those obtained previously using PC \#1 as the host PC. Our primary interest lies in observing the performance of each $S_{RAN}$ under varying CPU budgets, as highlighted in the previous section, where it was noted that not all CPU cores were fully utilized by any of the $S_{RAN}$ platforms. Please note that, PC \#1 operates at a base frequency of 3.5 GHz, while PC \#2 runs at a frequency of 2.9 GHz. The results are presented in Table~\ref{tab:HostPC}.

\begin{itemize}[wide, labelwidth=!, labelindent=0pt]

    \item \emph{Impact of the host PC on srsRAN:}
    There is a significant difference in the achieved performance of \textit{srsRAN} when changing the host PC. This effect is mainly evident in the \gls{DL} rate and the \gls{E2E} latency experienced by the \gls{UE}. When \textit{srsRAN} is running on PC \#2, the \gls{DL} rate drops to almost half of what is achievable when running \textit{srsRAN} on PC \#1 (from 92.1 Mbps to 46.9 Mbps). Additionally, the average latency increases by 37\%. These results are consistent with the findings reported in~\cite{haakegaard2024performance}, where the authors report achieved \gls{DL} and \gls{UL} rates of 123 Mbps and 39 Mbps, respectively, using a host PC more powerful than our PC \#1.
    
    \item \emph{Impact of the host PC on OAI-RAN:} \textit{OAI-RAN} appears to be less sensitive to the host PC, as the change in achieved performance when switching the host PC is not as significant as it is with \textit{srsRAN}. Although there is about a 9\% increase in the \gls{DL} rate, when using the more powerful PC, the \gls{UL} rate and average latency remain almost the same.
    
\end{itemize}



\begin{table}[h]
\centering
\caption{The effect of host PC computational power}
\label{tab:HostPC}
\resizebox{0.8\columnwidth}{!}{%
\begin{tabular}{|cc|cc|cc|}
\hline
\rowcolor[HTML]{329A9D} 
\multicolumn{2}{|c|}{\cellcolor[HTML]{329A9D}Performance Metric}  & \multicolumn{2}{c|}{\cellcolor[HTML]{329A9D}\begin{tabular}[c]{@{}c@{}}Data Rate\\ (in Mbps)\end{tabular}} & \multicolumn{2}{c|}{\cellcolor[HTML]{329A9D}\begin{tabular}[c]{@{}c@{}}Latency\\ (in ms)\end{tabular}} \\ \hline
\rowcolor[HTML]{68CBD0} 
\multicolumn{1}{|c|}{\cellcolor[HTML]{68CBD0}$S_{RAN}$} & Host PC & \multicolumn{1}{c|}{\cellcolor[HTML]{68CBD0}DL}                           & UL                             & \multicolumn{1}{c|}{\cellcolor[HTML]{68CBD0}Min}                         & AVG                         \\ \hline
\multicolumn{1}{|c|}{srsRAN}                            & PC \#1  & \multicolumn{1}{c|}{92.1}                                                 & 39.9                           & \multicolumn{1}{c|}{12}                                                  & 27                          \\ \hline
\multicolumn{1}{|c|}{srsRAN}                            & PC \#2  & \multicolumn{1}{c|}{46.9}                                                 & 39.9                           & \multicolumn{1}{c|}{11}                                                  & 37                          \\ \hline
\multicolumn{1}{|c|}{OAI-RAN}                           & PC \#1  & \multicolumn{1}{c|}{109}                                                  & 24.4                           & \multicolumn{1}{c|}{7.6}                                                 & 10.1                        \\ \hline
\multicolumn{1}{|c|}{OAI-RAN}                           & PC \#2  & \multicolumn{1}{c|}{99.4}                                                 & 24.5                           & \multicolumn{1}{c|}{7.1}                                                 & 10.5                        \\ \hline
\end{tabular}%
}
\end{table}

\section{Challenges \& Conclusion}~\label{sec:Conclusion}

In this paper, we presented one of the most comprehensive studies to date, on the performance achieved by 5G open-source software and \gls{COTS} hardware across 30 single-cell \gls{5G-SA} testbeds. We provided a precise nomenclature to characterize a 5G standalone testbed and a comprehensive set of metrics to assess performance. 
Our discussions into the performance of each testbed in both single and multi-\gls{UE} scenarios, highlighted how the type and location of each connected \gls{UE} impact performance. 
Additionally, we explored the interoperability of different UE types with various hardware and software elements of the \gls{RAN}. 
Finally, we evaluated the computational resource consumption of each software platform in both single and multi-\gls{UE} scenarios.

By defining three groups of locations, \textit{``good, fair, bad''}, for the connected \gls{UE}s, we first analyze how each 5G open-source \gls{RAN} platform performs, given good \gls{UE} positions. We then scatter the \gls{UE}s in different locations and observe the performance achieved in adverse conditions. Our findings indicate that if \textit{srsRAN} is executed on a powerful host PC, its performance can be superior given good \gls{UE} positions. However, the table turns when \gls{UE}s from further locations seek connection. In these scenarios, \textit{srsRAN} not only lacks the automatic power adjustment feature of \textit{OAI-RAN}, resulting in lower data rates for \gls{UE}s connected from distant locations, but also exhibits a discrepancy in the achieved performance based on the type of connected \gls{UE}. In this regard, \textit{OAI-RAN} being \gls{UE}-type agnostic and more robust, wins.
By analyzing the coverage support of four different testbeds, we revealed that the choice of $S_{RAN}$ is more influential than the choice of the \gls{SDR} device on coverage. Our results also showed that \textit{OAI-RAN} outperforms \textit{srsRAN}, in the \gls{E2E} latency.

One of the most critical aspects of this paper, as with any experimental study, is the clear definition of the test methodology.
This ensures that other researchers can hopefully reproduce our testbeds, and results uisng the elements we have selected and following our steps. 

\subsection*{Upcoming Challenges}

O-RAN, an industry-standard alliance and a significant player in the NG-RAN domain, has introduced a \gls{RAN} architecture that extends the one proposed by \gls{3GPP} with additional elements and interfaces. In their proposed architecture, they introduce three elements to split the 5G protocol stack: the Radio Unit (RU), Distributed unit (DU), and the Central Unit (CU), each responsible for running parts of the protocol stack. Currently, there is a growing interest in the research community not only to build \gls{5G-SA} experimental testbeds but also to develop testbeds that comply with the O-RAN standard.
At the time of this writing, very few O-RAN Radio Unit devices are available (e.g., the \textit{FlexFi O-RU} from LITE-ON Technology and Foxconn \textit{RPQN}). The main challenge in deploying an O-RAN compliant 5G-SA testbed is maintaining synchronization between the RU and DU, a process that requires precise timing and extensive communication. These challenges should be thoroughly addressed in future studies on O-RAN compliant, \gls{5G-SA} testbeds.



\printglossary[type=\acronymtype, title=Acronyms, toctitle=List of Acronyms]

\section*{Acknowledgment}

The authors would like to thank the kind volunteers who helped us during our measurement campaigns.

\ifCLASSOPTIONcaptionsoff
  \newpage
\fi


\printbibliography

@article{silveira2022tutorial,
  title={Tutorial on communication between access networks and the 5G core},
  author={Silveira, Lucas BD and de Resende, Henrique C and Both, Cristiano B and Marquez-Barja, Johann M and Silvestre, Bruno and Cardoso, Kleber V},
  journal={Computer Networks},
  volume={216},
  pages={109301},
  year={2022},
  publisher={Elsevier}
}

@inproceedings{mamushiane2023deploying,
  title={Deploying a Stable 5G SA Testbed Using srsRAN and Open5GS: UE Integration and Troubleshooting Towards Network Slicing},
  author={Mamushiane, Lusani and Lysko, Albert and Kobo, Hlabishi and Mwangama, Joyce},
  booktitle={2023 International Conference on Artificial Intelligence, Big Data, Computing and Data Communication Systems (icABCD)},
  year={2023},
  organization={IEEE}
}

@misc{alvesexperimental,
  title={Experimental comparison of 5G SDR platforms: srsRAN x OpenAirInterface},
  author={Alves, Ruan P and Alves, Jo{\~a}o Guilherme A da S and Camelo, Mikael R and de Feitosa, Wilker O and Monteiro, Victor F and Cavalcanti, Fco Rodrigo P}
}

@inproceedings{seidel2023get,
  title={How to Get Away with OpenAirInterface: A practical Guide to 5G RAN Configuration},
  author={Seidel, Mauri and Grohmann, Andreas Ingo and Sossalla, Peter and Kaltenberger, Florian and Fitzek, Frank HP},
  booktitle={2023 3rd International Conference on Electrical, Computer, Communications and Mechatronics Engineering (ICECCME)},
  year={2023},
  organization={IEEE}
}

@inproceedings{amini2024comparative,
  title={A Comparative Analysis of Open-source Software in an E2E 5G Standalone Platform},
  author={Amini, Maryam and Rosenberg, Catherine},
  booktitle={2024 IEEE Wireless Communications and Networking Conference (WCNC)},
  year={2024},
  organization={IEEE}
}

@inproceedings{chepkoech2023oss,
  title={Evaluation of OSS-Enabled OpenRAN Compliant 5G StandAlone Campus Networks},
  author={Chepkoech, Maurine and Modroiu, Elena-Ramona and Mwangama, Joyce and Corici, Marius and Magedanz, Thomas},
  booktitle={2023 International Conference on Electrical, Computer and Energy Technologies (ICECET)},
  year={2023},
  organization={IEEE}
}

@inproceedings{chepkoech2023evaluation,
  title={Evaluation of Open-Source Mobile Network Software Stacks: A Guide to Low-cost Deployment of 5G Testbeds},
  author={Chepkoech, Maurine and Mombeshora, Ngonidzashe and Malila, Bessie and Mwangama, Joyce},
  booktitle={2023 18th Wireless On-Demand Network Systems and Services Conference (WONS)},
  year={2023},
  organization={IEEE}
}

@inproceedings{amini20235G,
  title={5G DIY: Impact of Different Elements on the Performance of an E2E 5G Standalone Testbed},
  author={Amini, Maryam and El-Ashmawy, Ahmed and Rosenberg, Catherine and Amir, Khandani},
  booktitle={2023 IEEE Global Communications Conference (GLOBECOM)},
  year={2023},
  organization={IEEE}
}

@inproceedings{sahbafard2023performance,
  title={On the Performance of an Indoor Open-Source 5G Standalone Deployment},
  author={Sahbafard, Arash and Schmidt, Robert and Kaltenberger, Florian and Springer, Andreas and Bernhard, Hans-Peter},
  booktitle={2023 IEEE Wireless Communications and Networking Conference (WCNC)},
  year={2023},
  organization={IEEE}
}

@inproceedings{gabilondo20215g,
  title={5G SA Multi-Vendor Network Interoperability Assessment},
  author={Gabilondo, Alvaro and Fernandez, Zaloa and Mart{\'\i}n, {\'A}ngel and Viola, Roberto and Zorrilla, Mikel and Angueira, Pablo and Montalb{\'a}n, Jon},
  booktitle={2021 IEEE International Symposium on Broadband Multimedia Systems and Broadcasting (BMSB)},
  year={2021},
  organization={IEEE}
}

@inproceedings{amini2023implementing,
  title={Implementing an Open 5G Standalone Testbed: Challenges and Lessons Learnt},
  author={Amini, Maryam and El-Ashmawy, Ahmed and Rosenberg, Catherine},
  booktitle={IEEE INFOCOM 2023-IEEE Conference on Computer Communications Workshops (INFOCOM WKSHPS)},
  year={2023},
  organization={IEEE}
}

@techreport{TS38.401,
    author = {{3GPP}},
    day = {12},
    institution = {{3rd Generation Partnership Project (3GPP)}},
    month = {1},
    note = {Version 18.0.0},
    number = {38.401},
    title = {{NG-RAN; Architecture description}},
    type = {Technical Specification (TS)},
    url = {https://portal.3gpp.org/desktopmodules/Specifications/SpecificationDetails.aspx?specificationId=3219},
    year = {2024},}

@inproceedings{mubasier2023campus,
  title={Campus-Based Full-Scale and Portable Open-Source 5G SA Networks: Prototyping and Experiments},
  author={Mubasier, Kamar and Li, Frank Y and {\O}gaard, Jon Anders S and Vochin, Marius-Constantin},
  booktitle={2023 26th International Symposium on Wireless Personal Multimedia Communications (WPMC)},
  year={2023},
  organization={IEEE}
}

@misc{srsRAN,
    title = {{srsRAN Project}},
    howpublished = "\url{https://www.srsran.com}",
    note = "[Online; accessed 14-May-2024]",
}

@misc{OAI,
    title = {{OPEN AIR INTERFACE}},
    howpublished = "\url{https://openairinterface.org}",
    note = "[Online; accessed 14-May-2024]",
}

@misc{Open5GS,
    title = {{Open5GS}},
    howpublished = "\url{https://open5gs.org}",
    note = "[Online; accessed 14-May-2024]",
}

@misc{free5GC,
    title = {{free5GC}},
    howpublished = "\url{https://free5gc.org}",
    note = "[Online; accessed 14-May-2024]",
}

@misc{Lime,
    title = {{Lime microsystems}},
    howpublished = "\url{https://limemicro.com}",
    note = "[Online; accessed 14-May-2024]",
}

@misc{Nuand,
    title = {{Nuand}},
    howpublished = "\url{https://www.nuand.com}",
    note = "[Online; accessed 14-May-2024]",
}

@misc{Ettus,
    title = {{Ettus Research}},
    howpublished = "\url{https://www.ettus.com}",
    note = "[Online; accessed 14-May-2024]",
}

@article{haakegaard2024performance,
  title={Performance Evaluation of an Open Source Implementation of a 5G Standalone Platform},
  author={H{\aa}keg{\aa}rd, Jan Erik and Lundkvist, Henrik and Rauniyar, Ashish and Morris, Peter},
  journal={IEEE Access},
  volume={12},
  pages={25809--25819},
  year={2024},
  publisher={IEEE}
}

@misc{force5G,
    title = {{5G Switch - Force 5G Only}},
    howpublished = "\url{https://play.google.com/store/apps/details?id=com.sladjan.sava.petg&hl=en&gl=US}",
    note = "[Online; accessed 14-May-2024]",
}

@misc{sysmocom,
    title = {{sysmoISIM-SJA2 programmable SIM/USIM/ISIM cards}},
    howpublished = "\url{https://sysmocom.de/products/sim/sysmousim/index.html}",
    note = "[Online; accessed 14-May-2024]",
}

@inproceedings{bozis2024versatile,
  title={A Versatile 5G Standalone Testbed Based On Commodity Hardware},
  author={Bozis, Emmanouil-Zafeirios G and Sagias, Nikos C and Batistatos, Michael C and Kourtis, Michail-Alexandros and Xilouris, George K and Kourtis, Anastasios},
  booktitle={2024 Panhellenic Conference on Electronics \& Telecommunications (PACET)},
  year={2024},
  organization={IEEE}
}

@misc{iperf3,
  title        = {iperf3},
  howpublished = "\url{https://iperf.fr/}",
  note         = "[Online; accessed 14-May-2024]",
}

@misc{he.net,
  title        = {he.net-Network Tools},
  howpublished = "\url{https://play.google.com/store/apps/details?id=net.he.networktools&hl=en_CA&gl=US}",
  note         = "[Online; accessed 14-May-2024]",
}
%


%

\begin{IEEEbiography}[{\includegraphics[width=1in,height=1.25in,clip,keepaspectratio]{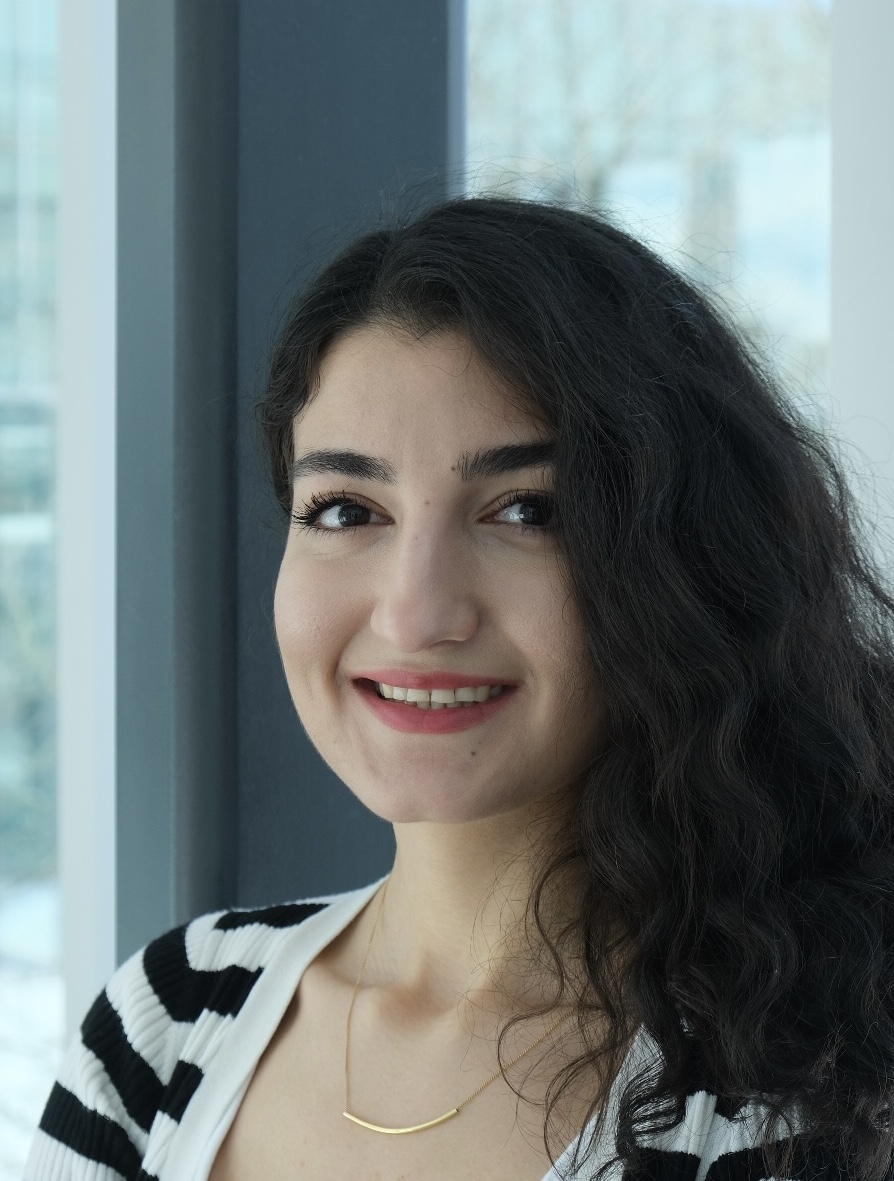}}]{Maryam Amini}
received her B.Sc., and M.Sc. degree from the Department of Computer Engineering at Iran University of Science and Technology in 2015 and 2017, respectively. Currently, she is pursuing her Ph.D. in the Department of Electrical and Computer Engineering at the University of Waterloo, Canada. Her research interests include Wireless Communications, Open RAN, and Experimental Testbeds.
\end{IEEEbiography}

\begin{IEEEbiography}[{\includegraphics[width=1in,height=1.25in,clip,keepaspectratio]{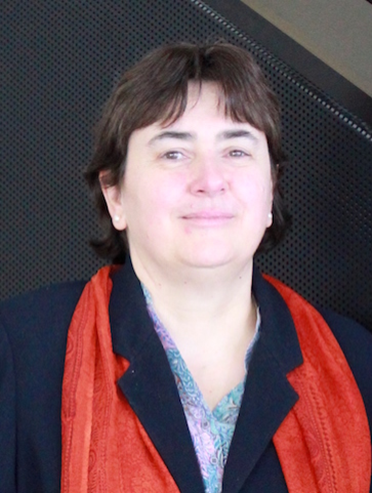}}]{Catherine Rosenberg}
(Fellow, IEEE) is currently a Professor with the Department of Electrical and Computer Engineering, University of Waterloo, ON, Canada.  She is also the Canada Research Chair in the Future Internet and the Cisco Research Chair in 5G Systems. Her research interests include networking and wireless. She is a Fellow of the Canadian Academy of Engineering. More information is available at  \url{https://uwaterloo.ca/scholar/cath}
\end{IEEEbiography}
\vfill



\end{document}